\documentclass[%
 reprint,
 amsmath,amssymb,
 aps,
]{revtex4-2}

\usepackage{graphicx}
\usepackage{dcolumn}
\usepackage{bm}
\usepackage{ulem}

\usepackage{xcolor}
\usepackage{wasysym}
\usepackage{relsize}

\begin{document}

\preprint{APS/123-QED}

\title{ Nonlinearity and Parametric Amplification of Superconducting Nanowire Resonators in Magnetic Field}

\author{M. Khalifa}
\author{J. Salfi}%
\affiliation{Stewart Blusson Quantum Matter Institute, University of British Columbia, Vancouver, BC, Canada.
}\affiliation{Department of Electrical and Computer Engineering, University of British Columbia, Vancouver, BC, Canada.
}




\date{\today}

\begin{abstract}
Nonlinear superconducting devices, typically based on Josephson Junction (JJ) nonlinearities, are the basis for superconducting quantum electronics, enabling, {\textit{e.g.}}, the formation of isolated two-level superconducting qubits and  amplifiers. While emerging spin, hybrid spin/superconducting (including Majorana), and nano-magneto-optical quantum systems could benefit tremendously from superconducting nonlinearities, the presence of strong magnetic fields in these systems are incompatible with conventional JJ devices, which are highly sensitive to applied magnetic fields. One potential solution is the use of kinetic inductance (KI) nonlinearity. To date, only linear kinetic inductance (KI) devices have been shown to operate in high magnetic fields,
while nonlinear KI device operation in high magnetic fields has received virtually no attention.
Here, we study the nonlinearity of superconducting nanowire (NW) KI resonators and their performance as parametric amplifiers for in-plane magnetic fields. We study the Kerr coefficients of NW KI resonators made from 10 nm-thin NbTiN films with characteristic impedance up to 3~k$\Omega$, and demonstrate both nondegenerate and degenerate parametric amplification, at magnetic fields up to 2 T, for the first time.
{\color{black}We find that narrow KI resonators of width 0.1 $\mu$m are robust, in terms of gain, dynamic range and noise, to magnetic fields up to $\sim$2 T, while wider KI resonators of width 1 $\mu$m suffer significant suppression in the gain around at fields well below $2$~T.}
Around 8 dB deamplification is observed for coherent states  for a 0.1 $\mu$m KI resonator, implying the capability of noise squeezing. 
These results open a new pathway to developing nonlinear quantum devices that operate in or generate high magnetic fields such as spin, hybrid spin/superconducting, and magneto-opto-mechanical devices.

\end{abstract}

\maketitle


\section{\label{sec:level1}Introduction}
Nonlinear superconducting devices are the basic components of superconducting quantum electronics. For example, inductive nonlinearities provide the anharmonicity essential for the superconducting circuits to operate as qubits \cite{krantz2019quantum, blais2021circuit}. The Josephson junction (JJ) in particular has been widely employed as the source of nonlinearity for superconducting qubits \cite{nakamura1999coherent, mooij1999josephson, van2000quantum, martinis2002rabi, yu2002coherent, vion2002manipulating, koch2007charge, manucharyan2009fluxonium, steffen2010high}. Additionally, Superconducting Quantum Intereference Devices (SQUIDs), which employ JJ nonlinearity, are used for making tunable couplers between qubits \cite{wallquist2006selective, osborn2007frequency, sandberg2008tuning, palacios2008tunable, niskanen2007quantum, chen2014qubit, mckay2016universal, weber2017coherent}, which provide high on-off ratio leading to lower gate error \cite{yan2018tunable, sete2021floating}, and they also enable parametric entangling gates which have high fidelity \cite{roth2017analysis, caldwell2018parametrically, reagor2018demonstration, sete2021parametric}. Such nonlinearity-based tunability has also been implemented in quantum memories for on-demand storing and transferring photons by tuning the coupling to a microwave resonator \cite{yin2013catch, pierre2014storage, sardashti2020voltage}.


One of the most important devices that exploits JJ nonlinearity is the Josephson parametric amplifier (JPA), which is widely used for near-quantum-limited measurements \cite{zimmer1967parametric, yurke1989observation, yamamoto2008flux, roy2015broadband, jeffrey2014fast}. JPAs can be used either in the degenerate (phase-sensitive) mode where they can in principle operate without adding any noise \cite{caves1982quantum, clerk2010introduction}, or in the nondegenerate (phase-preserving) mode where they ideally add half a photon of noise to the signal \cite{clerk2010introduction, caves2012quantum}. The former is used for squeezing coherent radiation or squeezing noise below the vacuum level (single mode squeezing) \cite{movshovich1990observation, castellanos2008amplification, zhong2013squeezing}, while the latter can be used for two-mode squeezing \cite{bergeal2010phase, eichler2011observation}. Applications of strongly squeezed states have recently extended beyond sensing to secure quantum communication \cite{pogorzalek2019secure, fedorov2021experimental} and fault tolerant quantum information processing \cite{fukui2018high}.

Recently, there has been an increasing interest in the kinetic inductance (KI) nonlinearity of superconducting nanowire (NW) devices made from NbTiN, NbN, and granular Al materials, a nonlinearity which is distributed along the NW \cite{maleeva2018circuit}, as an alternative to JJ-based devices. Nonlinearity distribution provide higher critical current, which can lead to a higher dynamic range and the ability to frequency-multiplex several devices \cite{eom2012wideband, vissers2016low}. Compared to JJ arrays in which the nonlinearity is also distributed over the whole array \cite{castellanos2007widely, castellanos2008amplification, eichler2014controlling, zhou2014high, planat2019understanding}, KI NWs naturally possess the diluted nonlinearity without any explicit junction formation and are relatively simple to fabricate.
In addition, higher-order nonlinearities have recently been identified as a limiting factor for gain, quantum efficiency, and squeezing in localized JJ-based devices \cite{kochetov2015higher, boutin2017effect}. The limitations of KI devices are less explored, however, recent work on KI devices has demonstrated nearly 50 dB of gain and 26 dB deamplification of coherent states\cite{parker2021near}.

Superconducting quantum electronic elements are rapidly finding applications in areas such as spin-based quantum systems  \cite{kobayashi2021engineering, schaal2020fast}, hybrid spin-photon systems \cite{mi2018coherent, borjans2020resonant, harvey2022coherent}, hybrid semiconductor-superconductor systems including Majorana zero modes \cite{mourik2012signatures,de2015realization}, and hybrid magneto-opto-mechanical systems \cite{zoepfl2020single}. What all of these systems have in common is magnetic fields that can easily reach 0.1~T to 1~T or higher. It is well known that even mT magnetic fields, such as the fields present near circulators in conventional setups \cite{lecocq2020microwave}, are detrimental to conventional JJ-based superconducting devices.  Therefore, magnetic-field resilient nonlinear superconducting devices would enable the integration of superconducting electronics with these novel quantum systems, allowing them to benefit from tunability, amplification, and squeezing provided by superconducting nonlinear devices.

\begin{figure*}
    \centering
    \includegraphics [width=0.585\textwidth] {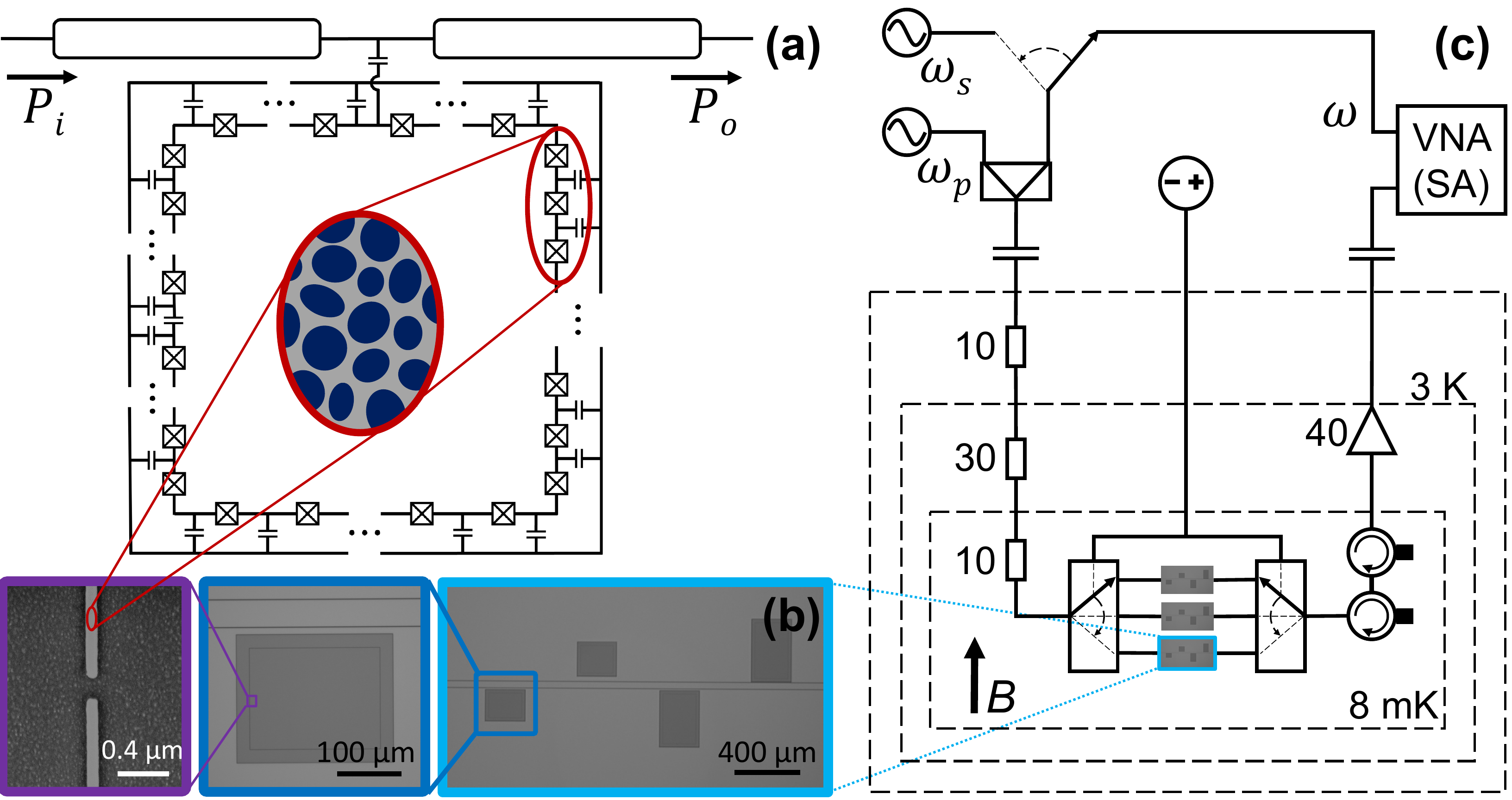}
    \includegraphics [width=0.35\textwidth] {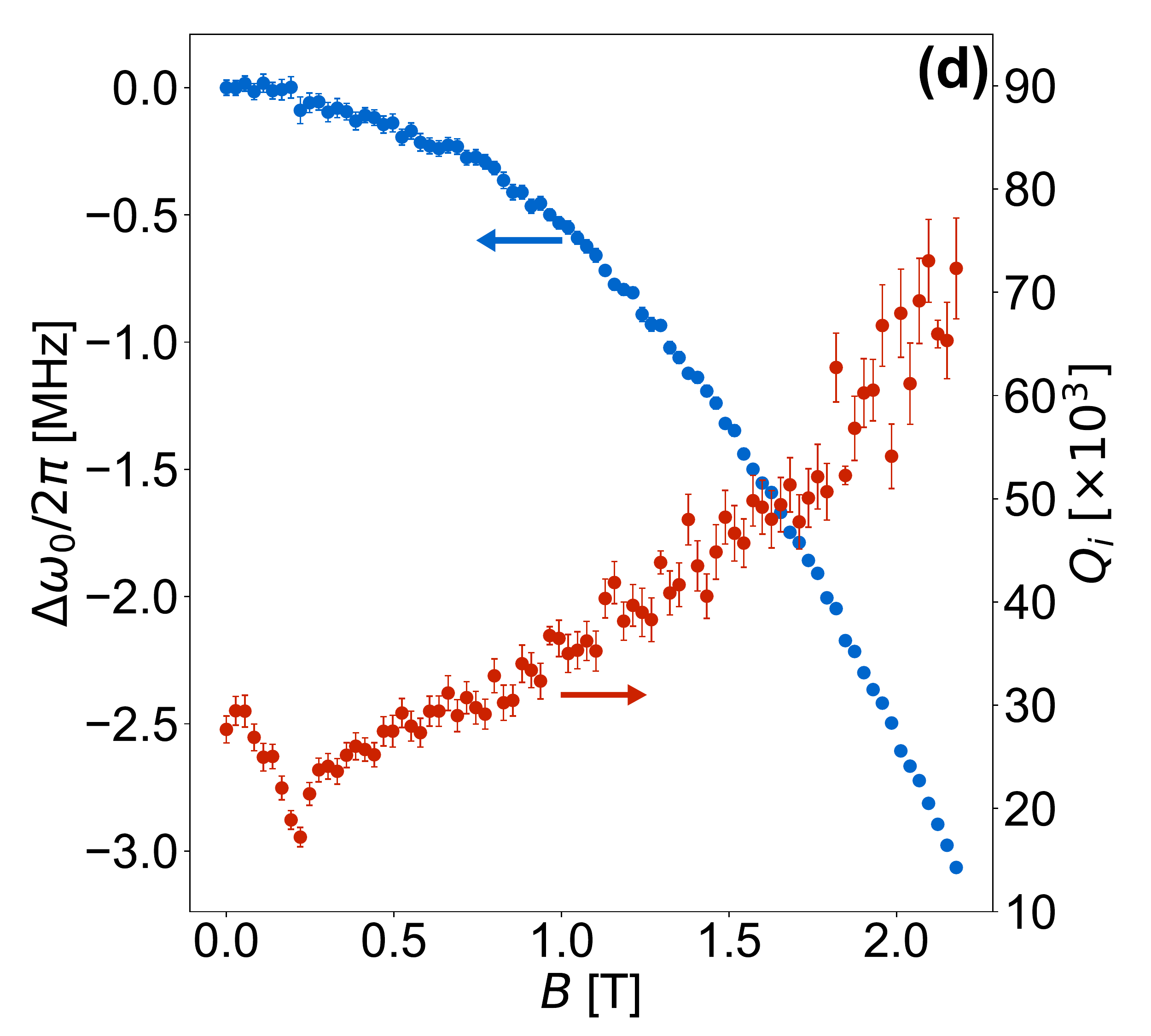}
    \caption{Device design and measurement. (a) Equivalent circuit diagram of the split-ring resonator in the hanger configuration. The nonlinear kinetic inductance is equivalent to an array of Josephson junctions and the split gap is represented by a capacitance between the terminals of the resonator. The resonator is capacitively coupled to the feedline. The inset is a cartoon for the highly disordered superconducting film. (b) SEM (left) and optical (middle and right) images showing the fabricated device on different scales. Several split-ring resonators are fabricated on each chip in the hanger configuration. The circuit diagram in (a) corresponds to the middle panel. (c) Schematic of measurement setup. Signals generated by VNA and/or microwave sources are combined and attenuated by 50 dB then sent to one of three chips placed at 8 mK. The output is amplified by a low-noise HEMT amplifier of 40 dB gain at 4 K and measured by VNA at room temperature. Short DC pulses are used to select the chip through latching microwave switches. (d) Resonance frequency and intrinsic quality factor versus magnetic field in the linear regime for the 0.1~$\mu$m wide KI NW resonator.}
    \label{fig:Overall}
\end{figure*}

Recently, magnetic field compatible nonlinearity has been explored in InAs-based Josephson Junctions \cite{pita2020gate} and granular Al NW junctions \cite{winkel2020implementation}, pushing the magnetic field range of operation up to 100s of mT. Although the resilience of linear KI resonator devices to magnetic fields is well established \cite{samkharadze2016high, kroll2019magnetic, yu2021magnetic}, the KI nonlinearity and devices based on it have yet to receive experimental attention at high magnetic fields. 

Here, we investigate, in magnetic fields as high as 3~T, the Kerr four-wave mixing (4WM) nonlinearity coefficient of NW KI resonators (without DC current bias) and their behaviour as degenerate and nondegenerate parametric amplifiers, for three different KI NW widths 0.1~$\mu$m, 0.3~$\mu$m, and 1~$\mu$m.  We find that increasing magnetic field causes an increase in internal quality factor and parabolic decrease in resonance frequency, consistent with previous reports \cite{zollitsch2019tuning}. For the 0.1 $\mu$m width, the quality factor increases from $30,000$ at $B=0$ to $70,000$ at 2~T. We also characterize the nonlinearity by measuring the self-Kerr coefficients in magnetic fields. The coefficients for the two 1~$\mu$m-wide wires exhibit a near-monotonic increase  with increasing magnetic field up to 3~T, by ~20\% and ~60\%, substantially larger than theoretical predictions. For the narrower wires, the dependence on magnetic field is not monotonic. Next, we demonstrate both nondegenerate and degenerate parametric amplification at 0 and 2~T, for the 0.1 $\mu$m width. Similar saturation gains are observed at $\sim$2~T and 0~T, but we find that the saturation gain is reached at a smaller pump power at 2~T, because of the combined effect of the significant increase in internal quality factor, and the small increase in Kerr coefficients, at 2~T.
{\color{black}For the 1~$\mu$m-wide resonators, the saturation gain decreases by 6 dB when magnetic field increased from 0 to 2 T.}

\section{\label{sec:level1} Methods}

Our resonators are made of a $t=10$ nm thick NbTiN superconducting film sputtered on a float-zone silicon substrate having $20$~k$\Omega\cdot$cm resistivity. The film is patterned by electron-beam lithography (EBL) into several split-ring NW resonators coupling capacitively to a CPW feedline of 50 $\Omega$ characteristic impedance, in the hanger configuration \cite{gao2008experimental, khalil2012analysis, megrant2012planar, samkharadze2016high, niepce2019high, basset2019high, yu2021magnetic}. The NbTiN film is etched by reactive ion etcher (RIE) in SF$_6$/O$_2$ plasma.

The nonlinear KI in such a nanowire superconducting film is often envisioned as the network of nanometer-scale superconducting islands that act as interconnected JJs. Figure \ref{fig:Overall}(a) depicts the equivalent circuit diagram of a single capacitively coupled split-ring resonator as an array of JJ shunted by capacitors. This circuit model accurately describes the dispersion relation and nonlinearity of high KI resonators \cite{maleeva2018circuit}. The nonlinearity (JJ-array like behaviour) arises from the strong disorder nature of the NbTiN film, which promotes the KI to be stronger than conventional magnetic inductance in our devices. Small junctions are formed between superconducting grains, as illustrated in the inset of Fig.\ref{fig:Overall}(a).

The critical temperature and sheet resistance of the film are measured to be $T_c=9.7$ K and $R_{\square} =340$ $\Omega/\square$, which gives sheet kinetic inductance $L_{\square} =48$ pH/$\square$ according to the relation $L_{\square} \approx \hbar R_{\square} / \pi \Delta_0$ \cite{annunziata2010tunable}, where $\Delta_0=1.764 k_B T_c$ is the superconducting energy gap at zero Kelvin and $k_B$ is Boltzmann constant. The critical current density of the film is measured to be $J_c=15$ mA/$\mu$m$^2$. 

Resonators of three widths 0.1, 0.3 and 1 $\mu$m are fabricated on three chips using the same source film. The estimated characteristic impedance of the three sets of resonators is $Z_r\approx3$, 1.5 and 0.7 k$\Omega$, respectively. The chips with the 0.1 $\mu$m and 1 $\mu$m resonators are etched at the same time, and the chip with the 0.3 $\mu$m resonators is etched separately. All chips are tested in a dilution refrigerator at 8 mK. The resonators are designed to have resonance frequencies in the range between 3 and 6 GHz, by varying their length $l_r$, and to have coupling quality factors between $10^3$ and $10^5$, by varying their distance from the feedline. A chip similar to the measured ones is shown in Fig.\ref{fig:Overall}(b) at three different scales. The transmission through the feedline is measured using a Vector Network Analyzer (VNA), whose signal is attenuated by total of 50 dB at the input of the feedline and the output is amplified by a low-noise high-electron-mobility transistor (HEMT) amplifier at the 4K stage, as shown by the measurement setup in Fig.\ref{fig:Overall}(c).




\section{Results and Discussion}
\subsection{Resonators Characterization in Linear Regime}

The measured transmission spectrum near resonance is fitted to the model in Refs. \cite{swenson2013operation, anferov2020millimeter, basset2019high}, which allows us to extract the low-power resonance frequency $\omega_{0}$, the intrinsic quality factor $Q_i$, the coupling quality factor $Q_c$ and the dimensionless power-dependent nonlinearity parameter $\xi$, as explained in Appendix \ref{app:NLResp}. 

The resonators are first characterized in the linear regime (i.e. $\xi \approx 0$) to determine the dominant source of intrinsic loss. At zero magnetic field, $Q_i$ exhibits the typical behavior for a resonator dominated by two-level-system (TLS) loss \cite{wang2009improving, sage2011study, sandberg2012etch}, with $Q_i \approx 4,000$ for unsaturated TLS and $30,000$ for totally saturated TLS for the 0.1 $\mu$m resonator (see appendix \ref{app:Losses}).

By ramping up the magnetic field we see two effects. Figure \ref{fig:Overall}(d) shows $\omega_0$ and $Q_i$ as a function of magnetic field $B$ at an excitation power that is high enough to saturate TLS but not too high to maintain the linear behavior. The resonance frequency decreases quadratically with the field strength according to $\Delta \omega_0/\omega_0 = -(\pi/48)[De^2t^2B^2/(\hbar k_B T_c)]$ \cite{samkharadze2016high, kroll2019magnetic, yu2021magnetic}, where $D$ is the electronic diffusion constant and $e$ is the electron charge constant. By fitting the data to this relation we get $D\approx 5$ cm$^2$s$^{-1}$, which is the same order of magnitude as the previously reported value for NbTiN \cite{samkharadze2016high}. The dip in $Q_i$ at $B \approx 0.2$ T, also observed before\cite{samkharadze2016high} is because of the coupling to the magnetic centers in the Si substrate. For magnetic fields above this dip, $Q_i$ increases with $B$ up to 2 T. A similar behavior has been reported before for narrow NbTiN resonators \cite{samkharadze2016high, zollitsch2019tuning}. This indicates that the resonator losses are dominated by the magnetic centers in the substrate even at zero magnetic field \cite{zollitsch2019tuning}, and that losses coming from magnetic vortices are negligible. The observed enhancement in the internal quality factor improves the performance of linear devices, and as we will see, can improve the performance of nonlinear KI devices in magnetic field. 

\subsection{Kerr Nonlinearity}

In our resonators, the dominant nonlinearity is 4WM (the Kerr effect). Three-wave mixing (3WM) is anticipated to be negligible because no DC current or flux is applied, so it is not explored. Different applications have different requirements on the strength of the nonlinearity. For example, in KI detectors the nonlinearity is a side effect and the Kerr coefficient should be as low as possible, while for qubits, large Kerr coefficients are preferable since they isolate the computation space,  \textit{e.g.}, Kerr coefficients of 100-400 MHz are typical for transmons \cite{blais2021circuit}. Parametric amplification falls between these two limits, where Kerr coefficient has to be higher than the nonlinear loss to maximize gain, but should be lower than the coupling loss to the feedline to maximize the dynamic range\cite{eichler2014controlling}, as discussed in next section.

The nonlinearity of the KI resonator was studied in detail in Ref. \cite{yurke2006performance}. The Hamiltonian of the resonator is
\begin{equation} \label{eqn:Hamil}
\begin{split}
    H = & \sum\limits_{n} \hbar \omega_n \left(a_n^\dagger a_n + \frac{1}{2}\right) + \sum\limits_{n} \frac{\hbar K_n}{2}  \left( a_n^\dagger a_n \right)^2 \\
    & + \sum\limits_{n,m \ne n} \frac{\hbar K_{mn}}{2} \left( a_m^\dagger a_m a_n^\dagger a_n \right)
\end{split}
\end{equation}
where $\omega_n$ and $K_n$ are the resonance frequency and self-Kerr coefficient of mode $n$, respectively, $K_{mn}$ is the cross-Kerr coefficient between modes $m$ and $n$, and $a_n$ ($a_n^\dagger$) is the annihilation (creation) operator of the intra-resonator field of mode $n$.

\begin{figure*}
    \centering
    \includegraphics [width=0.34\textwidth] {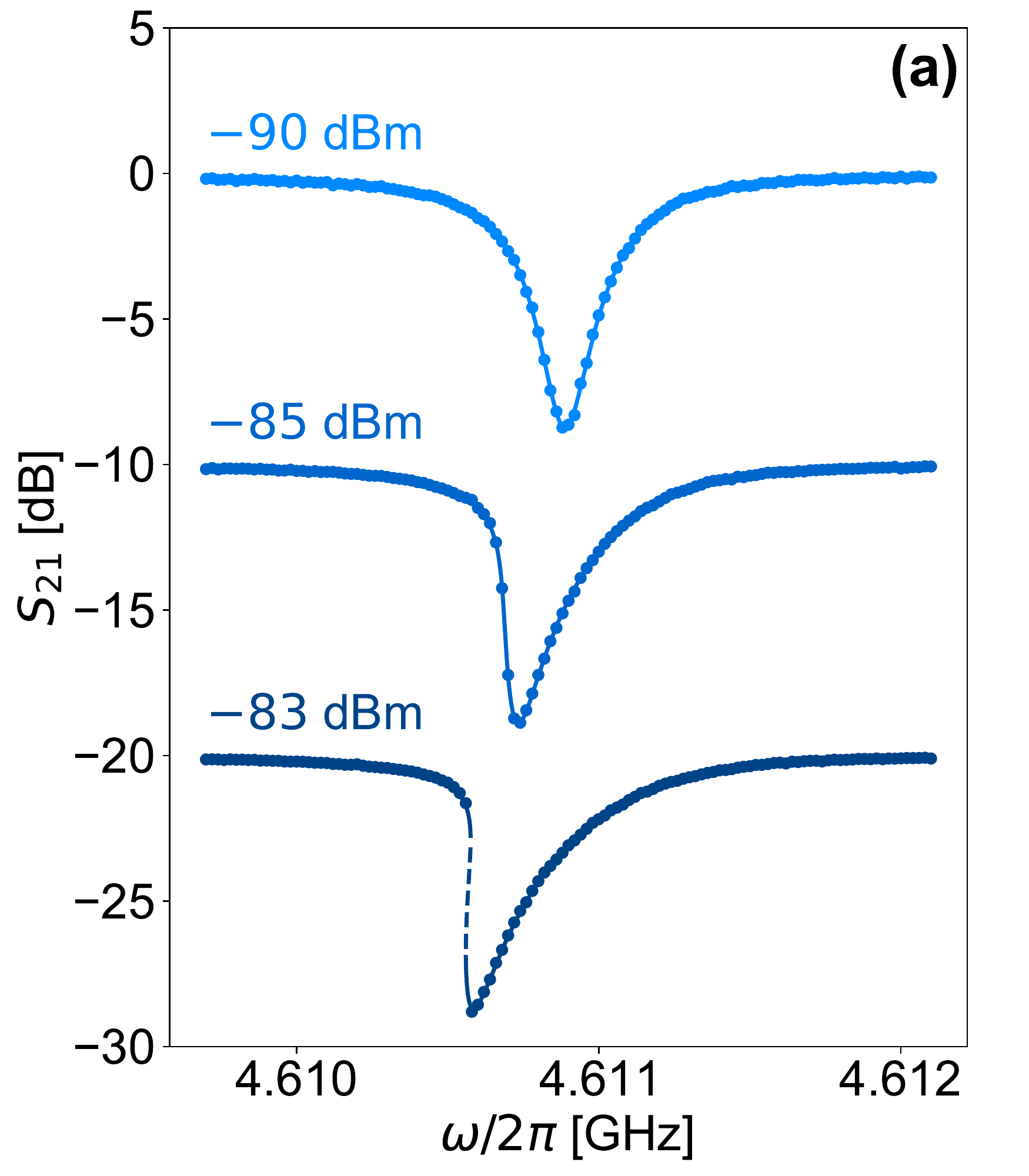}
    \includegraphics [width=0.29\textwidth] {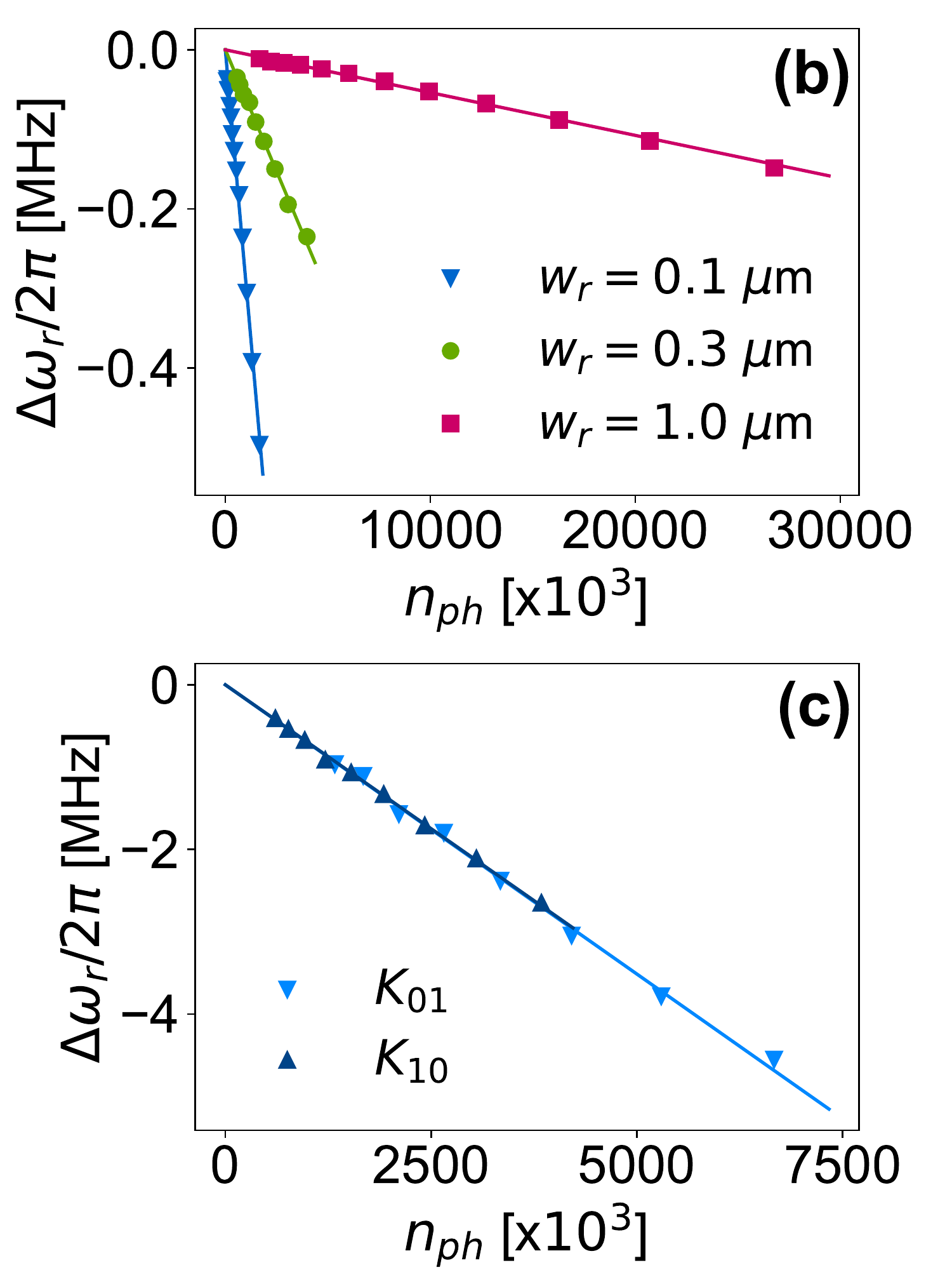}
    \includegraphics [width=0.34\textwidth] {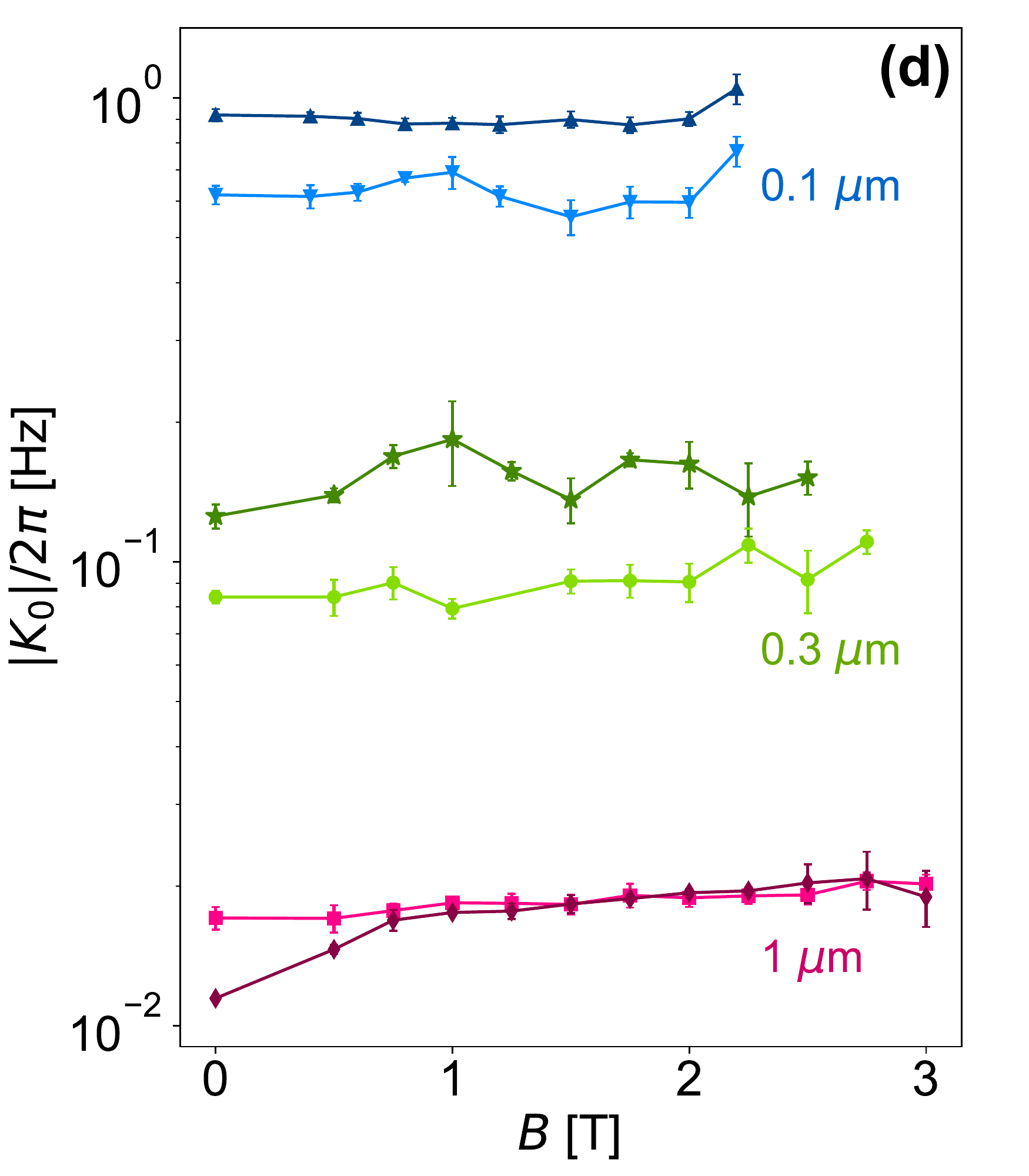}
    \caption{Kerr nonlinearity measurement. (a) $S_{21}$ measurements (dots) and fits (lines) for a resonator of 0.1 $\mu$m width at different powers; -90 dBm ($\xi=-0.03 \ll \xi_b$, negligible nonlinearity), -85 dBm ($\xi=-0.3 < \xi_b$, below bifurcation) and -83 dBm ($\xi=-0.46 > \xi_b$, above bifurcation). Lines are separated by 10 dB for clarity. The dashed part of the line at -83 dBm denotes the unstable states above bifurcation. (b) Shift in resonance frequency by varying input power versus estimated number of photons in the resonator for three resonators of different widths. Markers denote measurement points while solid lines are linear fits for extracting the self-Kerr coefficients: $K_0/2\pi = -0.584$, $-0.122$ and $-0.0108$ Hz for $w_r=0.1$, 0.3 and 1 $\mu$m, respectively. (c) Shift in resonance frequency $\omega_1/2\pi\approx 6.11$ GHz ($\omega_0/2\pi\approx 3.08$ GHz) of the first excited (fundamental) mode when pumping at the fundamental (first excited) mode versus estimated number of pump photons in the resonator for a resonator of width 0.1 $\mu$m. Markers denote measurement points while solid lines are linear fits for extracting the cross-Kerr coefficients: $K_{01}/2\pi = -1.406$ Hz and $K_{10}/2\pi = -1.398$ Hz. The pump power level at the first excited mode used for calculating $K_{10}$ is estimated relative to that at the fundamental mode from the relative values of the self-Kerr coefficients $K_0$ and $K_1$ (see text for details). (d) Extracted self-Kerr coefficients versus magnetic field for six resonators of widths 0.1, 0.3 and 1 $\mu$m, two of each width.}
    \label{fig:Nonlinear}
\end{figure*}

We characterize the nonlinearity of our resonators by experimentally measuring their self- and cross-Kerr coefficients. This requires estimating the average number of photons in the resonator corresponding to a certain driving power at the feedline using the extracted fitting parameters. Figure \ref{fig:Nonlinear}(a) displays the transmission response at three different powers. At low power ($\xi \approx 0$) the resonance frequency is $\omega_r = \omega_0$ and the resonator operates in the linear regime (top curve in Fig.\ref{fig:Nonlinear}(a)). By driving at a higher power and at a frequency closer to $\omega_r$, more photons couple to the resonator driving it into the nonlinear regime. As the average number of photons in the resonator increases, $\omega_r$ red-shifts resulting in the asymmetric resonance lineshape (middle curve in Fig.\ref{fig:Nonlinear}(a)). Above a certain power, corresponding to the onset of bifurcation ($\xi_b=-2/\sqrt{27}$ ignoring nonlinear loss \cite{anferov2020millimeter}), the resonator goes into the bistability regime, where the resonator has three valid states at the same frequency with different amplitudes; two of which are stable and one is unstable (bottom curve in Fig.\ref{fig:Nonlinear}(a)). This bistability feature has been used for digital threshold quantum detection in Josephson bifurcation amplifier, where the sensitivity is only limited by quantum fluctuations \cite{siddiqi2004rf, mallet2009single, vijay2009invited}.

The red-shift in $\omega_r$ at increasing power is proportional to the average number of photons in the resonator, $n_{\text{ph}}$, such that $\Delta \omega_r=K_0 n_\text{ph}/2$, where $K_0$ is self-Kerr coefficient depending on the critical current and kinetic inductance of the resonator. The value of $n_{\text{ph}}$ is determined for each input power as explained in Appendix \ref{app:NLResp}. We do the measurement at different powers and extract $K_0$ from the slope of the linear fit of $\Delta \omega_r$ versus $n_{\text{ph}}$. In this work, we focus on the trends with magnetic field and NW width, rather than the exact values of Kerr coefficients, so that uncertainties in $n_{\text{ph}}$, which is difficult to precisely determine, do not impact interpretation of the data.
Figure \ref{fig:Nonlinear}(b) presents $\Delta \omega_r$ versus $n_{\text{ph}}$ for three resonators of widths 0.1, 0.3 and 1 $\mu$m. The decrease in self-Kerr parameter $|K_{0}|$ as the resonator width $w_r$ increases comes from the dependence of $K_{0}$ on the resonator critical current $I_c=J_cw_rt$ and kinetic inductance $L_k=L_{\square}l_r/w_r$, where $|K_{0}| \propto \omega_0^2/L_k I_c^2$ \cite{yurke2006performance, anferov2020millimeter}.


Next, we extract the cross-Kerr coefficients which describe the strength of the intermode coupling in the resonator. This coupling can be used in circuit quantum electrodynamics for nondemolition measurements \cite{buks2006dephasing, helmer2009quantum, suchoi2010intermode, kumar2010exploiting}.
One approach is to measure the number of photons in a certain mode, called signal mode, by driving another mode, called detector mode, and detecting its dephasing \cite{buks2006dephasing, suchoi2010intermode} or frequency shift \cite{kumar2010exploiting}. The sensitivity of this scheme is determined by the magnitude of the cross-Kerr coefficient between the two modes \cite{buks2006dephasing}, where according to (\ref{eqn:Hamil}) the resonance shift of mode $n$ induced by pumping at mode $m$ is $\Delta \omega_r^{(n)} = K_{mn} n_\text{ph}^{(m)}/2$.
To extract $K_{mn}$ for $m\ne n$, we apply a two-tone technique  \cite{tancredi2013bifurcation, weissl2015kerr, krupko2018kerr, yu2021magnetic} by pumping mode $m$ with high power using a MW source ($\omega_p \sim \omega_m$) while probing mode $n$ with low power using the VNA, as depicted in Fig. \ref{fig:Overall}(c). We sweep the VNA frequency around $\omega_n$ for different pump powers at $\omega_m$. The shifts in the resonance frequency of the fundamental mode (i.e. $n=0$) of a resonator of width 0.1 $\mu$m when pumping at the first excited mode (i.e. $m=1$) and vice versa (i.e. $n=1$ and $m=0$) are shown in Fig. \ref{fig:Nonlinear}(c) as a function of the number of pump photons in the resonator.
A meaningful conclusion on the relation between $K_{01}$ and $K_{10}$, or between either of them and $K_0$, requires the power levels at $\omega_0$ and $\omega_1$ to be estimated accurately with respect to each other. The ratio between the self-Kerr coefficients can be used for this purpose since we know that $|K_n| \propto \omega_n^2$ \cite{yurke2006performance}. We take the power level at $\omega_0$ as a reference and calibrate the estimated power at $\omega_1$ such that $K_{1}/K_{0}=(\omega_1/\omega_0)^2=4$. The extracted values for $K_{01}$ and $K_{10}$ after power calibration are almost equal, as shown in Fig. \ref{fig:Nonlinear}(c), which agrees with previous theoretical predictions \cite{yurke2006performance}. We also find that the ratio $K_0/K_{01} \approx 0.4$, which is close to 0.375 calculated in \cite{maleeva2018circuit}.



The nonlinearity of the NW resonators is measured as a function of applied in-plane magnetic field (Fig. \ref{fig:Overall}(d)). The value of $|K_{0}|$ for two different 1~$\mu$m wide NW resonators monotonically increases by around $20\%$ and $60\%$ for magnetic fields up to $\sim 3$ T, the highest magnetic field we studied on those chips (Fig.~\ref{fig:Nonlinear}d, bottom). Notably, Kerr coefficients extracted from multiple 0.1 and 0.3 $\mu$m NW resonators do not exhibit a monotonic upward trend in $|K_0|$. Rather, $|K_0|$ increases for some values of $B$, and decreases in other regions, and exhibits a larger relative scatter in best-fit values compared to the 1 $\mu$m NWs (Fig.~\ref{fig:Nonlinear}(d), top and middle). Despite the presence of some scattering of data, the overall trend of for $|K_0|$ to increase between 0 and 2 T (2.5 T) for the 0.1 $\mu$m (0.3 $\mu$m) resonators. 

We compare the measured variation of the Kerr coefficient $K_0$ with magnetic field $B$ to the variation calculated from the theory of superconductivity. The Kerr coefficient depends on $T_c$, which decreases with increasing magnetic field due to breaking the time-reversal degeneracy of Cooper pairs \cite{tinkham2004introduction}. The relation between $K_0$ and $T_c$ originates from the dependence of $K_0$ on the superconducting parameters $I_c$ and $L_{\square}$, which both vary with $T_c$ \cite{annunziata2010tunable}, and it is given by (see Appendix \ref{app:KVsB})
\begin{equation}\label{eqn:Del_K}
    \frac{\Delta K_0}{K_0} = - \frac{\Delta T_c}{T_c} = \frac{\pi}{24} \frac{D e^2 t^2}{\hbar k_B T_c} B^2.
\end{equation}
According to this relation, the expected change in $K_0$ at $B=2$~T is only $\sim$0.1\%. Therefore, the observed increase in $|K_0|$ for the $1$~$\mu$m NWs (Fig.~\ref{fig:Nonlinear}(d), bottom) is more than 100 times larger than the theoretically predicted increase in $|K_0|$. It is difficult to pinpoint the cause for this enhancement compared to theoretical predictions. We speculate that it may reflect inhomogeneity present in the sputtered NbTiN film. For example, weak links, which may be viewed as sections of the NW with smaller critical currents and temperatures, could contribute to the enhancement of the Kerr coefficient. The less systematic variation in Kerr coefficients with magnetic field for the narrower wires may also be a signature of film or nanowire etching inhomogeneity.

\subsection{Nondegenerate Parametric Amplification}

\begin{figure*}
    \centering
    \begin{minipage}[b]{.6\linewidth}
        {\includegraphics [width=1\textwidth]{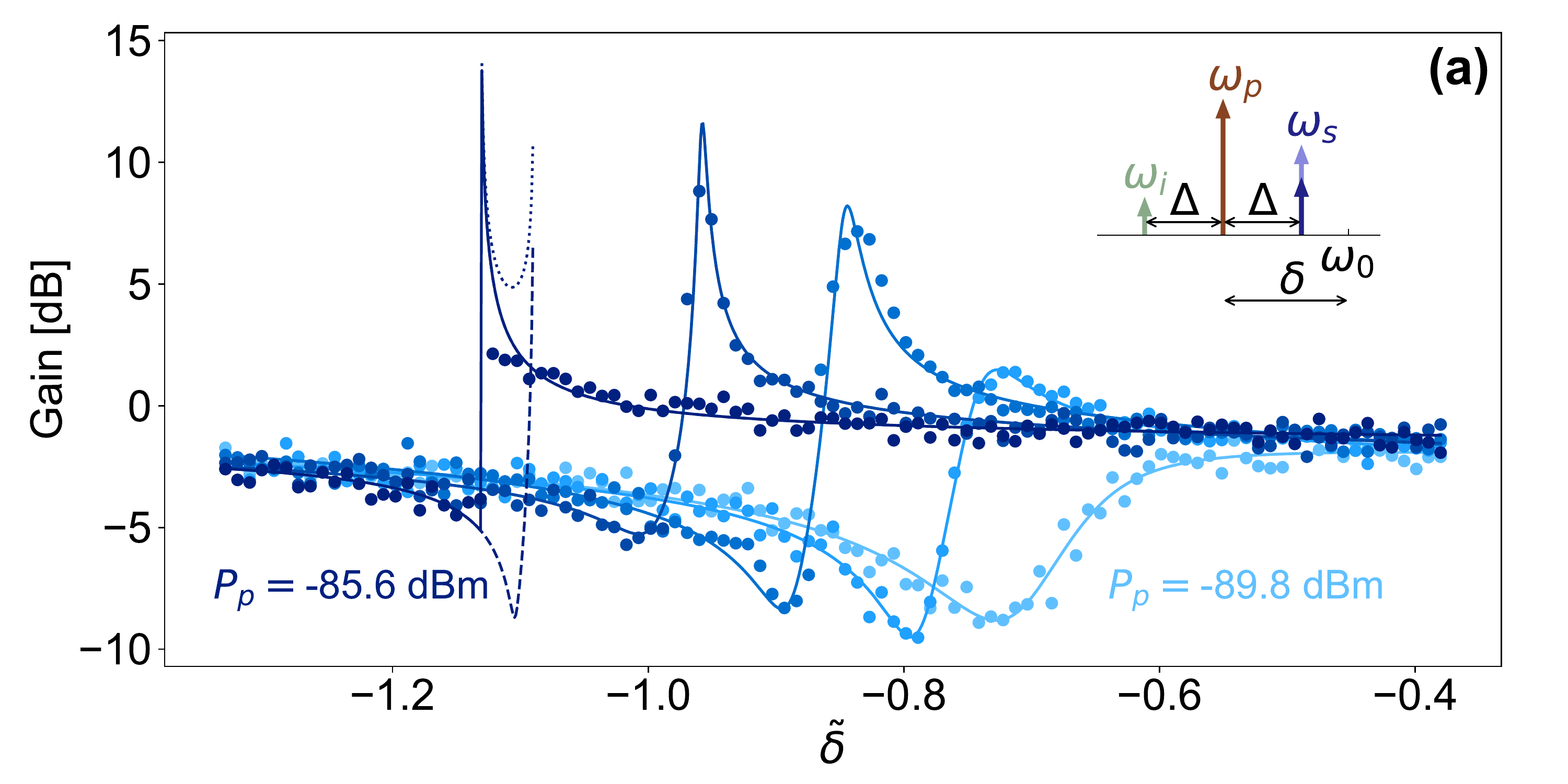}}
        {\includegraphics [width=1\textwidth]{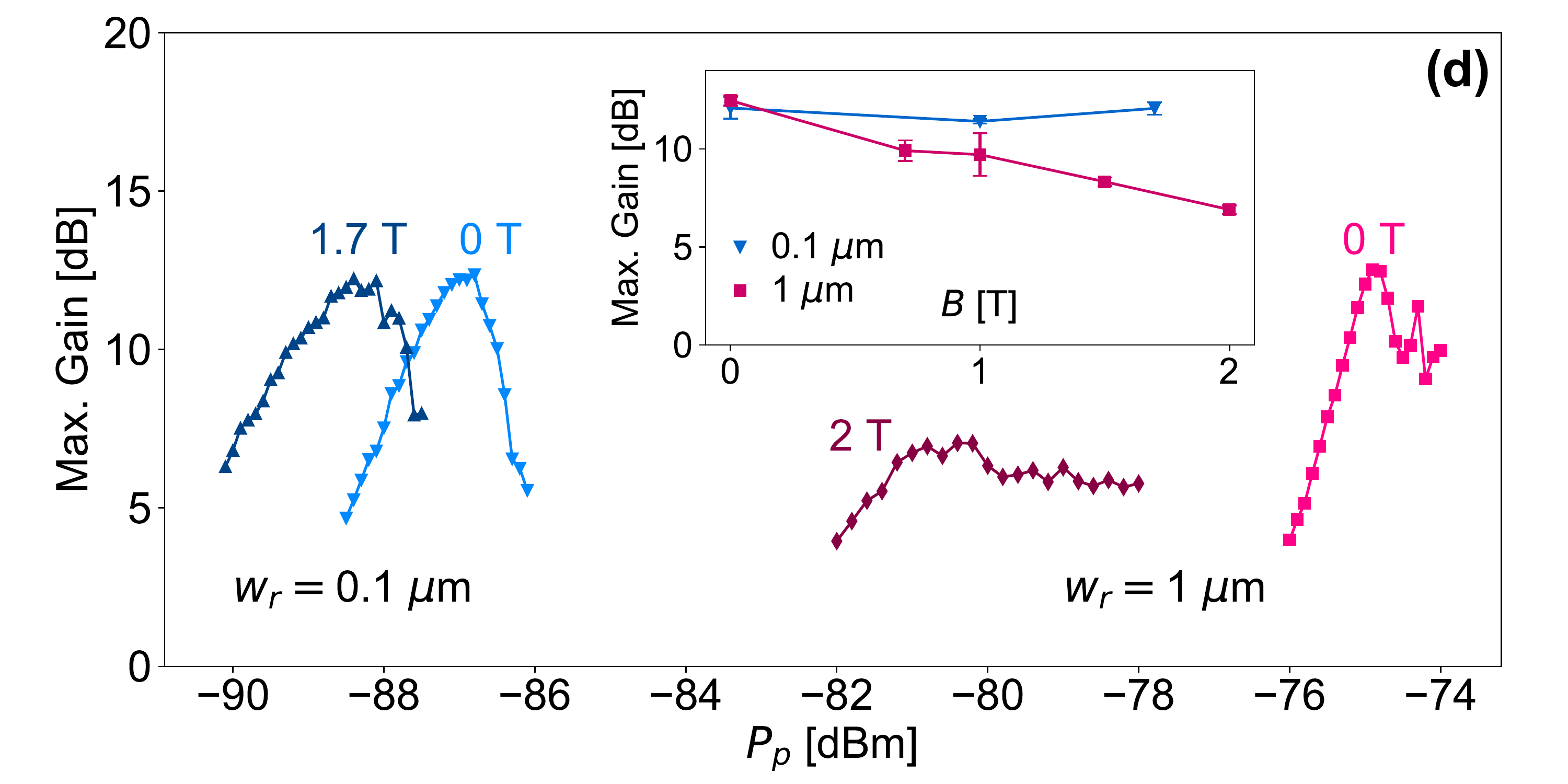}}
    \end{minipage}%
    \hfill
    \begin{minipage}[b]{.4\linewidth}
        {\includegraphics [width=1\textwidth] {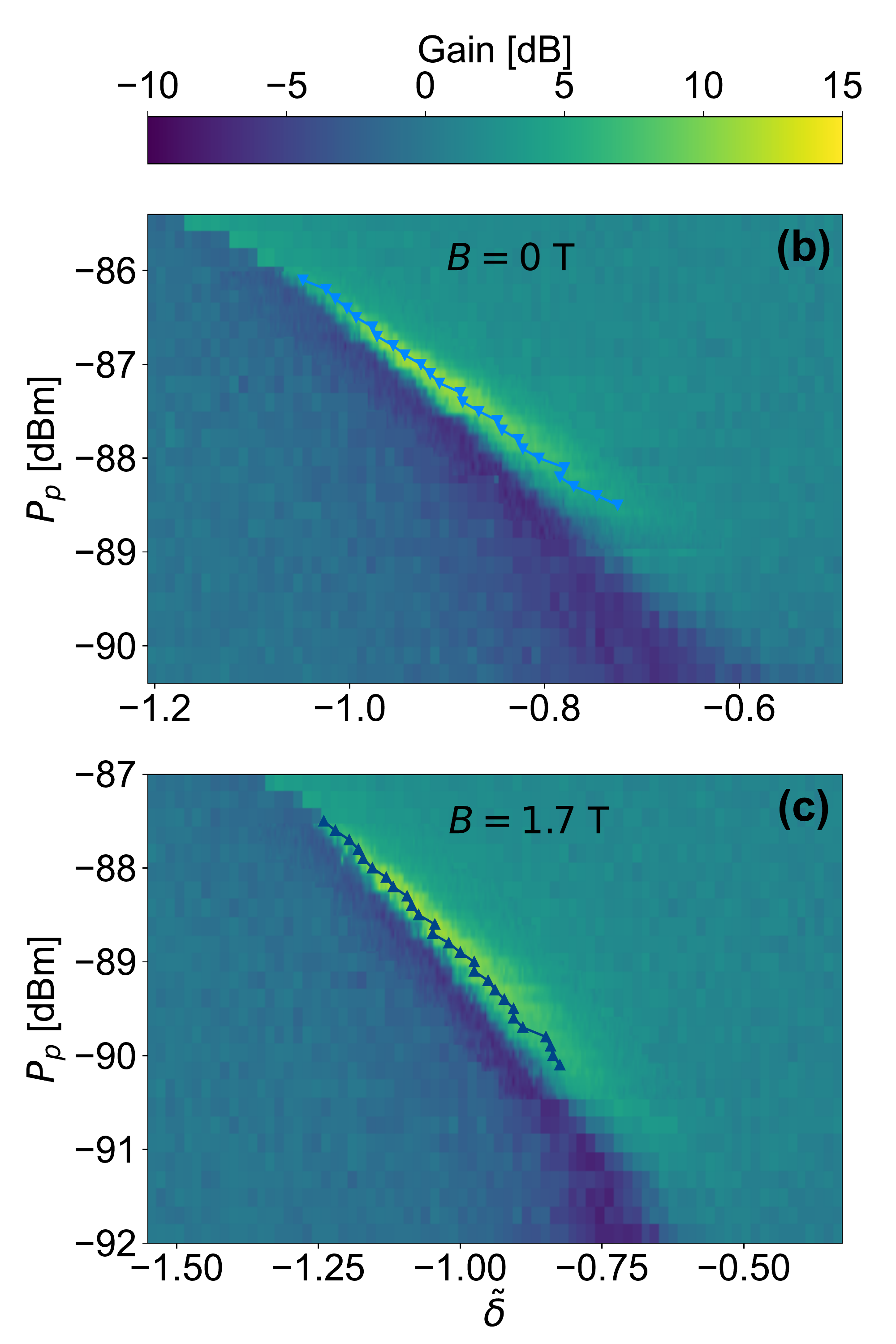}}
    \end{minipage}%
    \caption{Nondegenrate Parametric Amplification. (a) Measurement and fitting of signal gain versus reduced pump-resonance detuning, $\tilde{\delta}=(\omega_p-\omega_0) /(\kappa_c+\kappa_i)$, at different pump powers for a 0.1 $\mu$m-wide resonator. The highest-power line exhibits the bifurcation effect, where the solid part represents the fitting of the measured states, the dashed part represents the none-measured stable states, and the dotted part represents the unstable states.
    The inset shows the frequency configuration of the process (see text for details). (b), (c) Measured signal gain versus pump power, $P_p$, and reduced pump-resonance frequency detuning, $\tilde{\delta}$, at $B=0$ and 1.7 T, respectively, for the 0.1 $\mu$m resonator. The measurement is taken at downward pump frequency sweeping at each pump power. The marked-lines highlight the maximum gain for each pump power. (d) Maximum gain versus pump power for two resonators of widths 0.1 and 1 $\mu$m at zero and high and magnetic fields. The gain is maximized over the pump frequency. The two lines of the 0.1 $\mu$m resonator correspond to the lines highlighted in (b) and (c). The inset is the maximum gain over both pump frequency and power of the two resonators versus magnetic field. The solid lines are guide for the eye.}
    \label{fig:NDParaAmp}
\end{figure*}

A KI resonator can be operated as a parametric amplifier when pumped by a high-power tone. Here we consider a 4WM process, where the signal is slightly detuned from the pump within the linewidth of the same mode and the amplification occurs by the conversion of two pump photons into signal and idler photons. The performance of such parametric amplifiers was analyzed theoretically \cite{yurke2006performance} and measured experimentally \cite{tholen2007nonlinearities}. In the stiff-pump approximation, where the pump power at the output is assumed to be equal to the pump power at the input, the signal gain in the hanger configuration (Fig. ~\ref{fig:Overall}(a)) is given by \cite{anferov2020millimeter}
\begin{equation} \label{eqn:gs}
    g_s = 1 - \left(\frac{e^{i\phi}}{\cos \phi} \right) \left(\frac{\kappa_c}{\kappa_c+\kappa_i} \right) \frac{1/2 -i\left(\tilde{\delta}-2\xi n-\tilde{\Delta}\right)}{2\left(\lambda_+ + i\tilde{\Delta}\right)\left(\lambda_- + i\tilde{\Delta}\right)}
\end{equation}
where $\kappa_c = \omega_0/Q_c$ and $\kappa_i = \omega_0/Q_i$ are the coupling and intrinsic loss rates, respectively, $\xi$ is the reduced power-dependent nonlinearity parameter, $\eta$ is the reduced power-dependent nonlinear loss, $n$ is the reduced number of photons in the cavity, $\tilde{\delta}=(\omega_p - \omega_0)/(\kappa_c+\kappa_i)$ is the reduced pump-resonance detuning, $\tilde{\Delta}=(\omega_s - \omega_p)/(\kappa_c+\kappa_i)$ is the reduced signal-pump detuning, $\phi$ is a phase representing the asymmetry in the feedline and $\lambda_{\pm} = 1/2 \pm \sqrt{(\xi n)^2+ - (\tilde{\delta}-2\xi n)^2}$.

We measure the performance of the KI resonators as nondegenerate (phase insensitive) parametric amplifiers in magnetic field. We stimulate a 4WM process by supplying a pump tone detuned from the resonance frequency by $\delta = \omega_p-\omega_0$ and a low-power signal detuned from the pump tone by $\Delta = \omega_s-\omega_p$, where $|\Delta| \ll \kappa_c+\kappa_i$. Inside the resonator two pump photons at $\omega_p$ are converted into two photons at $\omega_s$ and $2\omega_p-\omega_s$, leading to the amplification of the signal and the generation of a new tone (idler) at frequency $\omega_i = 2\omega_p-\omega_s$, as illustrated in the inset of Fig. \ref{fig:NDParaAmp}(a). The amplified output power of the signal is then measured by a spectrum analyzer using a measurement bandwidth well below $|\Delta|$.
We define the signal gain at $\omega_s=\omega_p+\Delta$ as the ratio of the signal output power at $\omega_s$ when pump is on to the signal output power at $\omega'_s$ when pump is off, where $\omega'_s$ is far from resonance. The signal output power at $\omega'_s$ when pump is off is used as an approximation for the signal input power at $\omega_s$ (the signal output power at $\omega_s$ when pump is off does not reflect the signal input power at $\omega_s$ when $\omega_s$ is close to the resonance frequency).

The gain data for $\Delta=10$ kHz for a resonator of width 0.1 $\mu$m at $B=0$ T is fitted with the expression in (\ref{eqn:gs}) and plotted in Fig. \ref{fig:NDParaAmp}(a) at different pump powers. The gain increases with pump power until the onset of bifurcation where the system enters the bistability regime. The parametric amplifier exhibits large gain when operated at pump power and frequency near the bifurcation point. The condition for large amplification is that the system can access the bifurcation regime, which is satisfied when the nonlinear (i.e. two-photon) loss rate is smaller than $|K_0|/\sqrt{3}$  \cite{yurke2006performance}.



\begin{figure}
    \centering
    \includegraphics [width=1\linewidth] {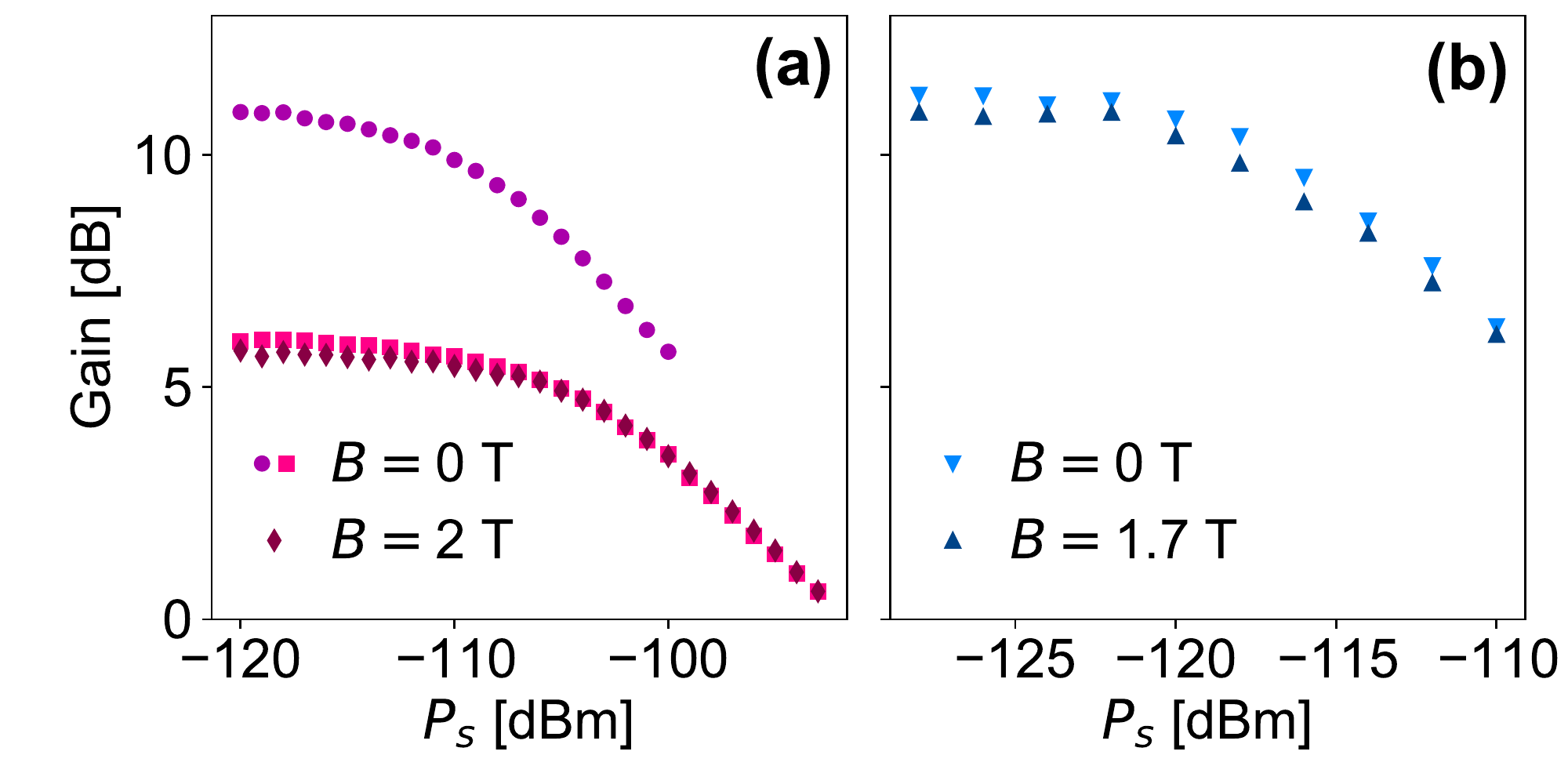}
    \includegraphics [width=1\linewidth] {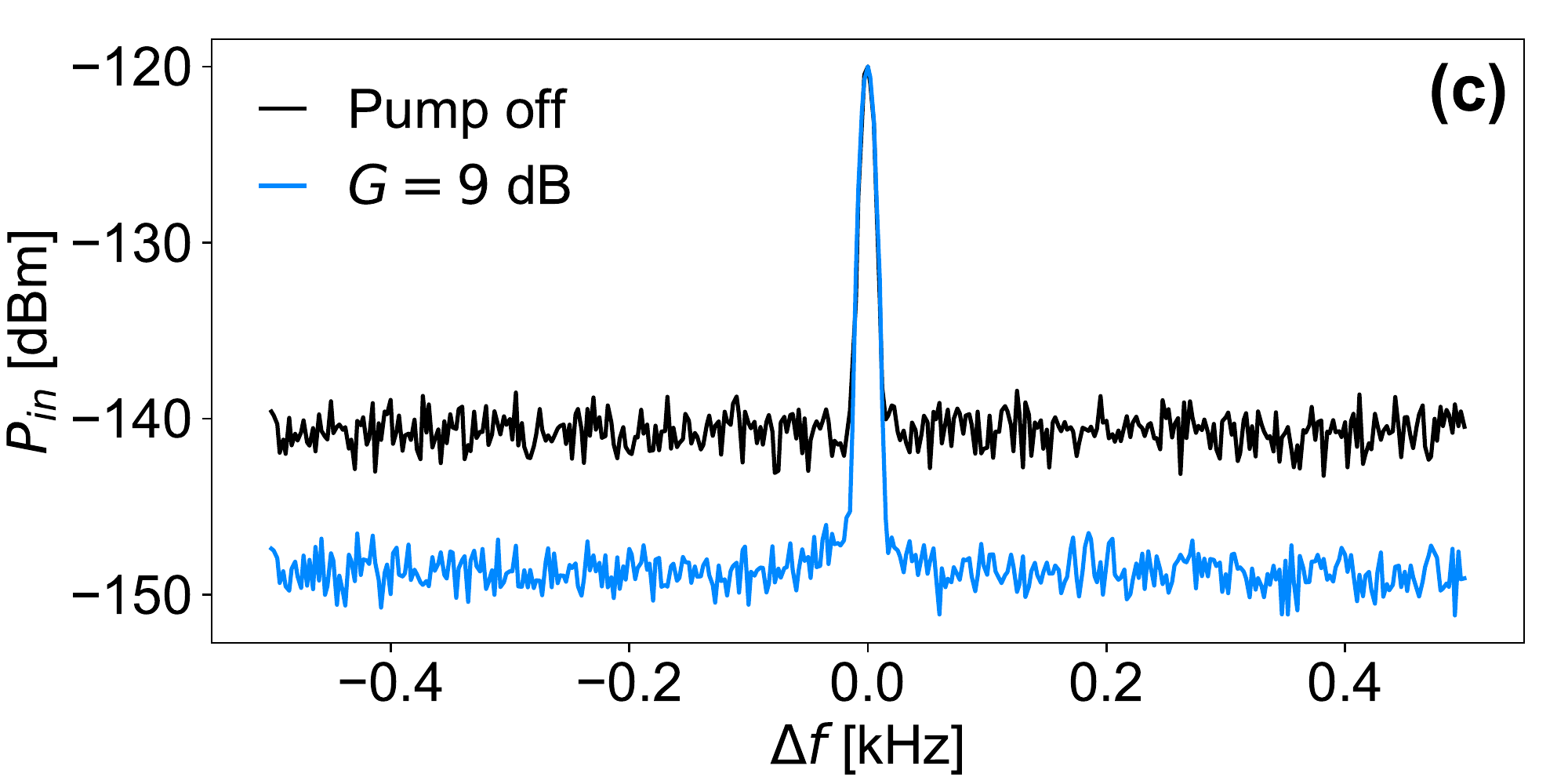}
    \includegraphics [width=1\linewidth] {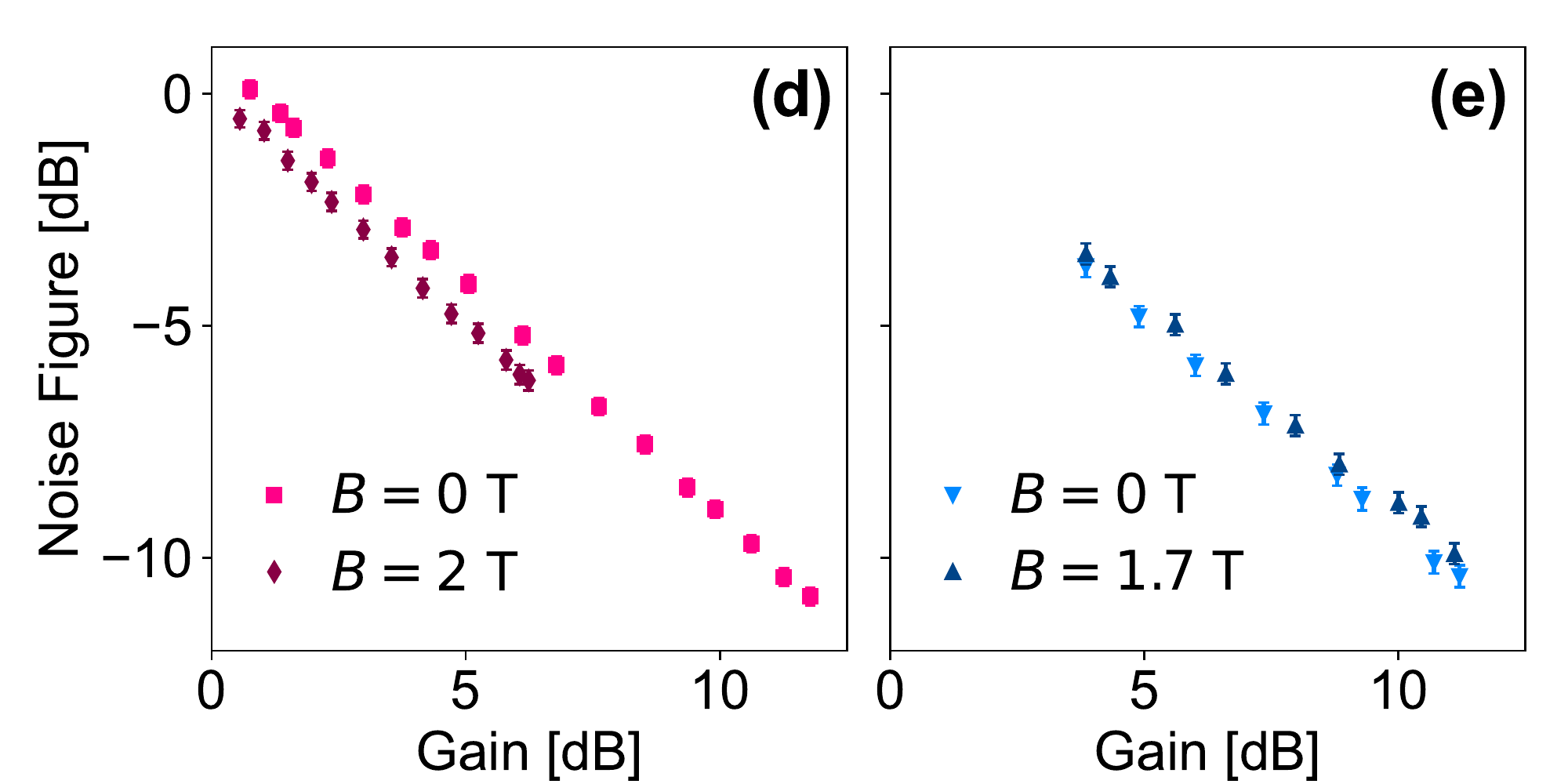}
    \caption{Parametric Amplifier Characterization. (a), (b) Gain as a function of signal power $P_s$ for the 1 $\mu$m and the 0.1 $\mu$m resonators, respectively. In (a), the data denoted by {\smaller$\CIRCLE$} and {\smaller$\blacksquare$} at $B=0$ T correspond to different pump powers ($P_p =-75.3$ and $-75.8$ dBm, respectively), hence different gain values. (c) Noise power referred to the input of the 0.1 $\mu$m KI PA at $B=0$ T in the presence of a coherent tone when pump is on at 9 dB gain and when it is off. (d), (e) Noise figure as a function of gain for the 1 $\mu$m and the 0.1 $\mu$m KI PAs, respectively, at zero and high magnetic fields.}
    \label{fig:DRNF}
\end{figure}

\begin{figure*}
    \centering
    \includegraphics [width=0.195\textwidth] {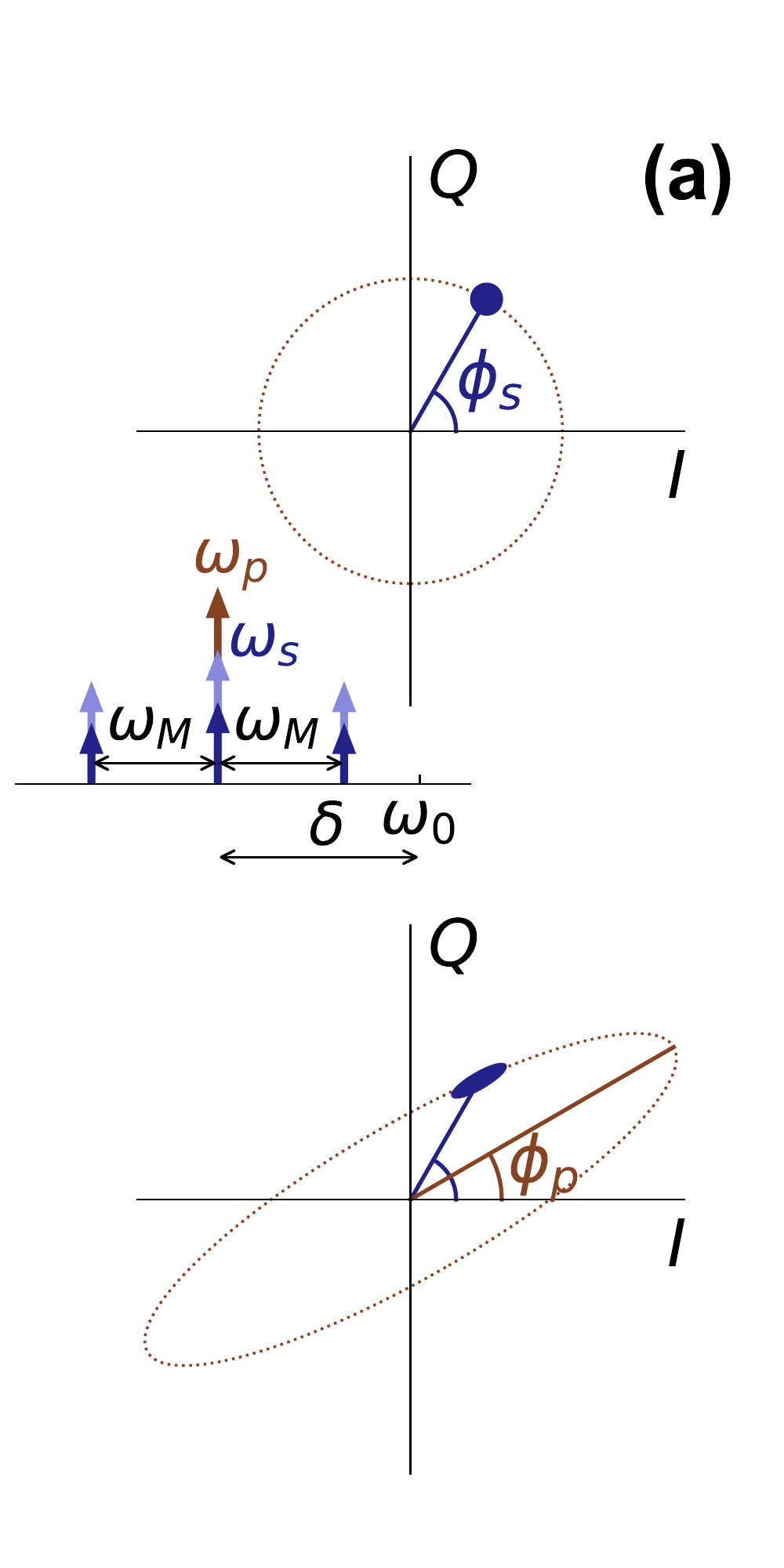}
    \includegraphics [width=0.39\textwidth] {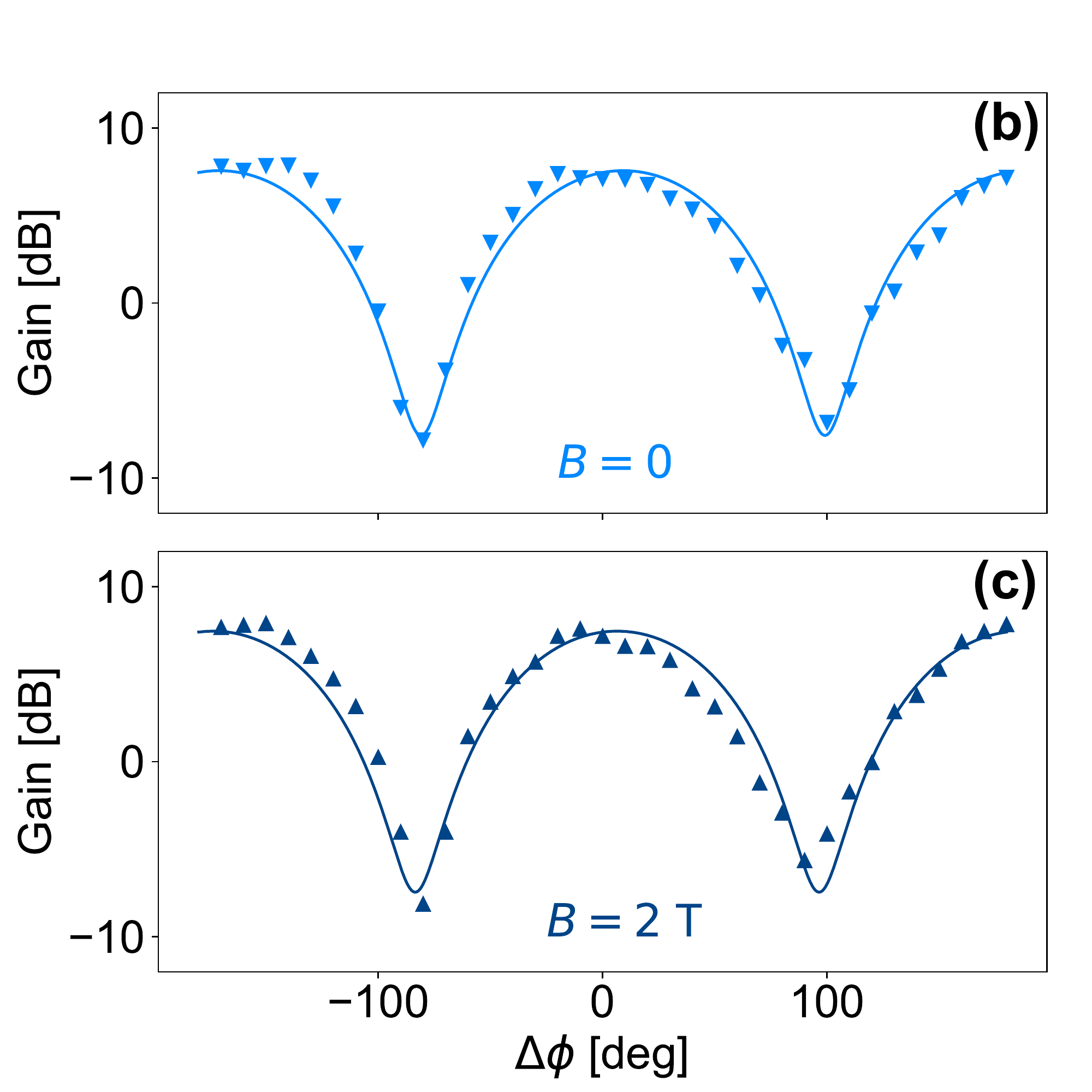}
    \includegraphics [width=0.39\textwidth] {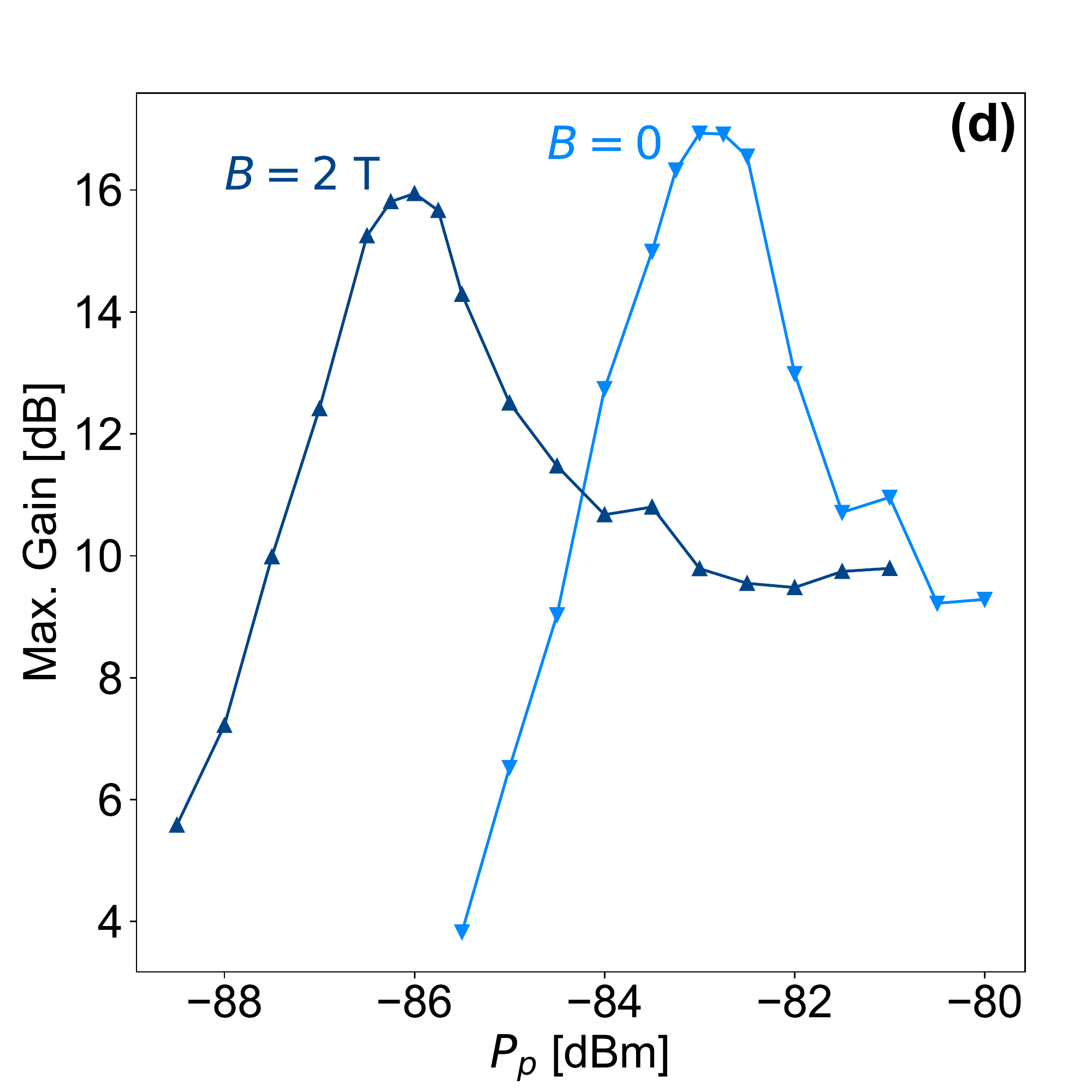}
    \caption{Degenerate Parametric Amplification. (a) Schematic of the output signal phase space without (top panel) and with (bottom panel) pump tone. The former is a coherent state whose amplitude is independent of phase, while the later is a squeezed state whose amplitude depends on its phase relative to the pump phase. The inset shows the frequency configuration of the experiment. The signal is amplitude-modulated at $\omega_M$ and the gain is measured at the modulation sideband. (b),(c) Measured signal gain as a function of the relative phase between the signal and pump tones at  $B=0$ and $2$ T, respectively. The points at which gain is above (below) zero indicate the amplification (deamplification). (d) Maximum gain versus pump power. The gain is maximized over phase and frequency at each power value. The solid lines are guide for the eye.}
    \label{fig:DParaAmp}
\end{figure*}

Figures \ref{fig:NDParaAmp}(b) and (c) show the full measurement of the gain versus pump power and pump frequency for the same resonator of width 0.1 $\mu$m at 0 and 1.7~T, respectively. In both cases,  gain is observed in the NW resonator. The marked points which are connected by solid lines in Fig.~\ref{fig:NDParaAmp}(b) and (c) indicate the frequency where the maximum gain is obtained for each pump power. The maximum gain is plotted in Fig. \ref{fig:NDParaAmp}(d) versus the pump power for the $0.1$~$\mu$m and $1$~$\mu$m NW resonators. At zero magnetic field, the maximum gains of the two resonators are almost equal, $\sim$12.5 dB. However, to reach this gain, the 1 $\mu$m resonator requires higher pump power due its lower Kerr coefficient.
This verifies that the maximum gain is insensitive to the Kerr nonlinearity, so that high gains can still be achieved for resonators with arbitrarily small Kerr-coefficients if pumped by sufficiently high power. However, as mentioned before, pumping with high power is not desirable since it heats up the resonator by generating thermal quasi-particles. Above a certain threshold, the quasi-particle losses will dominate the total losses of the resonator decreasing the intrinsic quality factor and increasing the noise level.
{\color{black}On the other hand, the gain of the two resonators is quite different at high magnetic field. The maximum gain of the 0.1 $\mu$m resonator is almost not affected at 1.7 T, except for a small shift in the power at which it is observed, while the maximum gain of the 1 $\mu$m is significantly suppressed at 2 T.
The measured maximum gain is plotted versus magnetic field in the inset of Fig. \ref{fig:NDParaAmp}(d). The 0.1 $\mu$m resonator exhibits a stable gain over the whole measurement range of $B$, which is limited by the microwave switches in our setup, while the gain of the 1 $\mu$m resonator degrades gradually with $B$. To confirm this result, another 1 $\mu$m resonator was measured and similar degradation for its gain versus $B$ was observed (not shown).}
The shift in the power at which the maximum gain is observed at high field, (see Fig. \ref{fig:NDParaAmp}(d) main panel), is due to the fact that the resonators have higher $Q_i$ and slightly higher $|K_0|$ in magnetic field, as illustrated in Fig. \ref{fig:Overall}(d) and Fig. \ref{fig:Nonlinear}(d), respectively. The higher $Q_i$ translates into a larger average number of photons in the resonator given a certain pump power, while the higher $|K_0|$ translates into a larger $\xi$ given a certain number of photons in the resonator.
{\color{black}Most importantly, these results demonstrate the sensitivity of the gain levels obtained at high magnetic field to the resonator width, where narrower KI devices can maintain their high gains up to higher values of magnetic field.}

For operating points near the bifurcation, gain can additionally be limited by pump depletion, where as the signal output power approaches the pump power, the gain begins to saturate. This defines the dynamic range of the device and it is limited by the pump power at which the bifurcation occurs. For a lossless device ($\kappa_i \ll \kappa_c$), the bifurcation power is inversely proportional to $|K_0| Q_c^2$ (see Appendix \ref{app:NLResp}). Therefore lower $|K_0|$ and $Q_c$ gives larger dynamic range. However, this requires operating at higher pump power leading to the excitation of thermal quasi-particles and lowering $Q_i$ \cite{eichler2014controlling}.
{\color{black}We determine the dynamic range of our devices by measuring their gain as a function of input signal power $P_s$, and observing the signal power at which 1-dB compression of the gain occurs. In Fig. \ref{fig:DRNF}(a) we show the data for the 1 $\mu$m resonator for two different gains of 11 and 6 dB, determined by pump power and frequency, at 0 T, and for a gain of 6 dB at 2 T. The 11 dB gain gives lower 1-dB compression point, \textit{i.e.} smaller dynamic range, $\sim-110$ dBm compared to $-105$ dBm for the 6 dB gain, which agrees with both theory and experiment in literature \cite{eichler2014controlling, parker2021near}. The presence of magnetic field does not affect the 1-dB compression point as long as the device operates at a similar gain, as shown in Fig. \ref{fig:DRNF}(a) for the case of 6 dB gain with and without field. The 0.1 $\mu$m KI PA also exhibits a dynamic range that is insensitive to magnetic field, where the 1-dB compression point occurs at $\sim-118$ dBm at both 0 and 1.7 T, as shown in Fig. \ref{fig:DRNF}(b) for a gain of 11 dB. The smaller dynamic range of the 0.1 $\mu$m KI PA compared to the 1 $\mu$m one is attributed to the larger Kerr coefficient of the 0.1 $\mu$m resonator.}

{\color{black}We also characterize the noise properties of our KI PAs. Figure \ref{fig:DRNF}(c) shows the noise power referred to the input of the 0.1 $\mu$m KI PA in the presence of a coherent input signal at $\sim-120$ dBm. The input referred noise decreases by roughly the same amount of the gain when the resonator is pumped for amplification, improving the signal-to-noise ratio (SNR) by the same factor. Similar results are obtained at high magnetic field and for the 1 $\mu$m resonator.
In addition, we measure the noise figure as a function of gain. We define the noise figure as the ratio between the SNR when pump is off, SNR$_\text{off}$, and the SNR when pump is on at a certain gain, SNR$_\text{on}(G)$, where in this definition the SNR$_\text{off}$ is used as an indication of the SNR at the input of the KI PA. The results for the 1 $\mu$m and the 0.1 $\mu$m KI PAs are shown in Figs. \ref{fig:DRNF}(d) and (e), respectively, at zero and high magnetic fields. The noise figure decreases linearly with gain for both KI PAs over the whole range of amplification, even with the existence of the magnetic field.
This means that the gains of these resonators are not high enough to reach the saturation point of SNR$_\text{on}(G)$, where the noise at the mixing chamber, either from quantum fluctuations or added noise by the KI PA, dominates over the noise added by the HEMT or thermal noise at higher stages. Despite this limitation in our devices, we can conclude that the magnetic field did not significantly enhance the added noise of the 0.1 $\mu$m KI PA up to 12 dB gain.}

{\color{black}These results show that, other than the gain suppression of the 1 $\mu$m resonator, the high magnetic field does not affect the characteristics of our KI PAs. Therefore, KI devices of sufficiently narrow widths, $\sim$0.1 $\mu$m, are capable of operating efficiently as PAs at high magnetic fields up to $\sim$2 T.}

\subsection{Degenerate Parametric Amplification}

Finally we explore the phase-sensitive amplification of our NW resonators in magnetic field. Here the device operates in the degenerate mode, where the pump and signal tones are set to the same frequency ($\Delta = 0$), so the idler is generated at the same frequency. This results in squeezing the output phase space, with squeezing level and direction determined by the power and phase of the pump tone, respectively, as illustrated by the schematic in Fig. \ref{fig:DParaAmp}(a). The magnitude of the output signal is then either amplified or deamplified depending on its phase relative to the pump phase. Similarly the variance of the output signal undergoes the same effect, which can be used for squeezing thermal and vacuum noise. Squeezing level up to 26 dB has been recently achieved using a KI-based degenerate parametric amplifier \cite{parker2021near} operating in zero magnetic field.

In order to probe the gain of our NW resonators in the degenerate regime, we amplitude-modulate the signal with a sinusoidal wave of small frequency $\omega_M$ and measure the gain at the sideband with measurement bandwidth much less than $\omega_M$. We use $\omega_M=20$ kHz, and we adjust the pump frequency for maximum gain. Similar to the case of nondegenerate amplification, the gain here is obtained by comparing the output sideband power when the pump is on to that when the pump is off at a reference point far from resonance. Figures \ref{fig:DParaAmp}(b) and (c) show the measured gain versus $\Delta \phi = \phi_s-\phi_p$ for a resonator of width 0.1 $\mu$m at 0 and 2 T with estimated pump power at the resonator $-81$ and $-83$ dBm, respectively. In both figures the gain oscillates at phase period $\pi$, such that the gain varies from a maximum value of 7.5 dB at $\Delta \phi=0$ to a minimum value of -7.5 dB at $\Delta \phi = \pm\pi/2$.
In Fig. \ref{fig:DParaAmp}(d) the maximum gain per phase difference and pump frequency is plotted as a function of pump power at 0 and 2 T. The gain peaks at the bifurcation power with almost the same value of $\sim16$ dB at 0 and 2 T. Similar to the nondegenerate case, there is a power shift between the measurements at 0 and 2 T, due to the combination of higher $Q_i$ and slightly higher $|K_0|$ at 2~T.
These results demonstrate the ability of KI resonators to operate as phase-sensitive parametric amplifiers at high magnetic field without compromising their amplification or squeezing capabilities.

\section{Conclusion}
We experimentally investigate the ability of superconducting NW KI resonators to function as nonlinear devices in high magnetic fields for the first time. By characterizing the Kerr nonlinear coefficients of the resonators, we show that their value generally increases for in-plane magnetic fields up to 3 T. The Kerr coefficients of the 1 $\mu$m wide NW resonators increase by more than 100 times more than anticipated. We speculate that this could be due to inhomogeneities in the 10 nm thick NbTiN film used to build the devices, which are revealed through the application of magnetic fields on the NW; a topic that deserves further study. By stimulating a 4WM process, we operate the KI resonators as parametric amplifiers in high magnetic field, and measure their phase-preserving and phase-sensitive gains. {\color{black}Phase-preserving gains above 12 dB are obtained for KI resonators of widths 0.1 and 1 $\mu$m at 0 T, limited by our device design. At magnetic fields around 2 T, the gain of the 1 $\mu$m KI PA decreases by about 5 dB, while the gain of the 0.1 $\mu$m maintains its value. Both the 1 $\mu$m and the 0.1 $\mu$m KI PAs exhibit robust performance in terms of dynamic range and noise figure around 2 T.}
Phase-sensitive gain of 16 dB is obtained at 2 T for the 0.1 $\mu$m KI PA. We also observe signal deamplification at 2 T, which implies the potential of these devices for noise squeezing. These results show that KI-based nonlinear devices, including parametric amplifiers, can function efficiently at high magnetic field. This would enable systems requiring high magnetic field for operation, such as spin and majorana based quantum systems, to benefit from the wide range of applications that superconducting nonlinear devices can offer. Parametric amplification in particular could significantly enhance the qubit measurement fidelity in these quantum systems.

\begin{acknowledgments}
This work was undertaken with support from the Stewart Blusson Quantum Matter Institute (SBQMI), the National Science and Research Council of Canada through the Discovery Grant scheme, the Canadian Foundation for Innovation through the John Edwards Leaders Foundation scheme, and the Canada First Research Excellence Fund, Quantum Materials and Future Technologies Program. MK acknowledges financial support from the SBQMI QuEST fellowship program. MK and JS acknowledge Shabir Barzanjeh and Tim Duty for helpful suggestions and feedback on the manuscript. The authors acknowledge CMC Microsystems for the provision of computer aided design tools that were essential to obtain the results presented here. This research was supported in part through computational resources provided by Advanced Research Computing at the University of British Columbia.
\end{acknowledgments}

\appendix

\section{Nonlinear Response} \label{app:NLResp}
The transmission spectrum near the resonance frequency of a resonator of resonance frequency $\omega_0$, coupling loss rate $\kappa_c=\omega_0/Q_c$ with $Q_c$ as the coupling quality factor to the feedline, instrinsic loss rate $\kappa_i=\omega_0/Q_i$ with $Q_i$ as the intrinsic quality factor, and Kerr parameter $K_0$, can generally be described by the nonlinear Diameter Correction Method (DCM) using the transfer function \cite{swenson2013operation, anferov2020millimeter}
\begin{equation}\label{eqn:S21}
    S_{21} = 1 - \frac{e^{i\phi}}{\cos \phi} \frac{Q_r}{Q_c} \frac{1}{1+j2\tilde{\delta}_r}
\end{equation}
where $Q_r=Q_iQ_c/(Q_i+Q_c)$ is the total quality factor of the resonator, $\phi$ is a phase representing the asymmetry in the transmission line, and $\tilde{\delta}_r=(\omega-\omega_r)/(\kappa_c+\kappa_i)$ is a dimensionless parameter representing the detuning between the signal frequency $\omega$ and the resonator resonance frequency $\omega_r$, reduced by normalizing to the resonator total loss, $\kappa_c+\kappa_i$. In the linear regime, $\omega_r$ is independent of the signal power and frequency ($\omega_r=\omega_0$, where $\omega_0$ is the resonance frequency in the low-power resonance frequency). As the power increases and $\omega$ gets closer to $\omega_r$, more photons couple to the resonator and it goes into the nonlinear regime where $\omega_r$ changes by an amount $\Delta \omega_r=\omega_r - \omega_0 < 0$. This shift follows from the nonlinear kinetic inductance of the resonator, and it results in a shift in the detuning parameter given by \cite{zmuidzinas2012superconducting, swenson2013operation}
\begin{equation}\label{eqn:delx}
    \tilde{\delta}_r - \tilde{\delta}_0 \approx \frac{-\Delta \omega_r} {\kappa_c+\kappa_i} = - \frac{2\xi}{1+4\tilde{\delta \mkern 0mu}_r^2}
\end{equation}
where $\tilde{\delta}_0=(\omega-\omega_0)/(\kappa_c+\kappa_i)$ is the reduced detuning in the linear regime, and $\xi$ is the dimensionless power-dependent nonlinearity parameter defined as the product of the reduced Kerr coefficient, the reduced coupling loss rate and the reduced incident photon rate, such that \cite{basset2019high}
\begin{equation}\label{eqn:xi}
    \xi = \left(\frac{K_0}{\kappa_c+\kappa_i} \right) \left(\frac{\kappa_c}{\kappa_c+\kappa_i} \right) \left(\frac{P / \hbar \omega_0}{\kappa_c+\kappa_i} \right)
\end{equation}
where $P$ is the input power at the feedline. The parameters $\omega_0$, $Q_i$, $Q_c$, $\phi$ and $\xi$ are extracted by fitting transmission measurement to $S_{21}$ in (\ref{eqn:S21}).
An important step in the fitting process is to solve for $\tilde{\delta}_r$ from (\ref{eqn:delx}), which is a cubic equation of $\tilde{\delta}_r$ and has three roots. 
Below the onset of bifurcation, only one solution is physical (has a real root for $\tilde{\delta}_r$). At high power, the three solutions could be valid and the system could enter the bistability regime. In that case, the three solutions correspond to three states of the resonator, two of which are stable as explained in the main text. This regime is only accessible above the onset of bifurcation which occurs at the critical values of $\xi$ and $\tilde{\delta}_0$; $\xi_b = -2/\sqrt{27}$ and $\tilde{\delta}_{0,b}=-\sqrt{3}/2$ \cite{anferov2020millimeter,eichler2014controlling}. At this point, the system has three degenerate states with $\tilde{\delta}_{r,b} = \tilde{\delta}_{0,b}/3=-1/2\sqrt{3}$.
Therefore, from (\ref{eqn:delx}), the shift in resonance frequency at the onset of bifurcation is $\Delta\omega_{r,b} = -(\kappa_c+\kappa_i)/ \sqrt{3}$.

The average number of photons stored in the resonator, $n_{\text{ph}}$, is directly related to the resonance frequency shift by $\Delta\omega_r=n_{\text{ph}}K_0/2$. From this relation along with (\ref{eqn:delx}) and (\ref{eqn:xi}), $n_{\text{ph}}$ can be calculated from $P$ as
\begin{equation}\label{eqn:n_ph}
    n_{\text{ph}} = -(\tilde{\delta}_r - \tilde{\delta}_0) \frac{\kappa_c+\kappa_i}{K_0/2}
    = \frac{4}{1+4\tilde{\delta \mkern 0mu}_r^2} \frac{\kappa_c (P/\hbar\omega_0)}{(\kappa_c+\kappa_i)^2}.
\end{equation}
Also, the number of photons at the onset of bifurcation can be directly calculated from $\Delta \omega_{r,b}$ as $n_{\text{ph},b} = 2(\kappa_c+\kappa_i) / \sqrt{3} |K_0|$.




\section{Power Dependence of Losses}
\label{app:Losses}

\begin{figure}
    \centering
    \includegraphics [width=\linewidth] {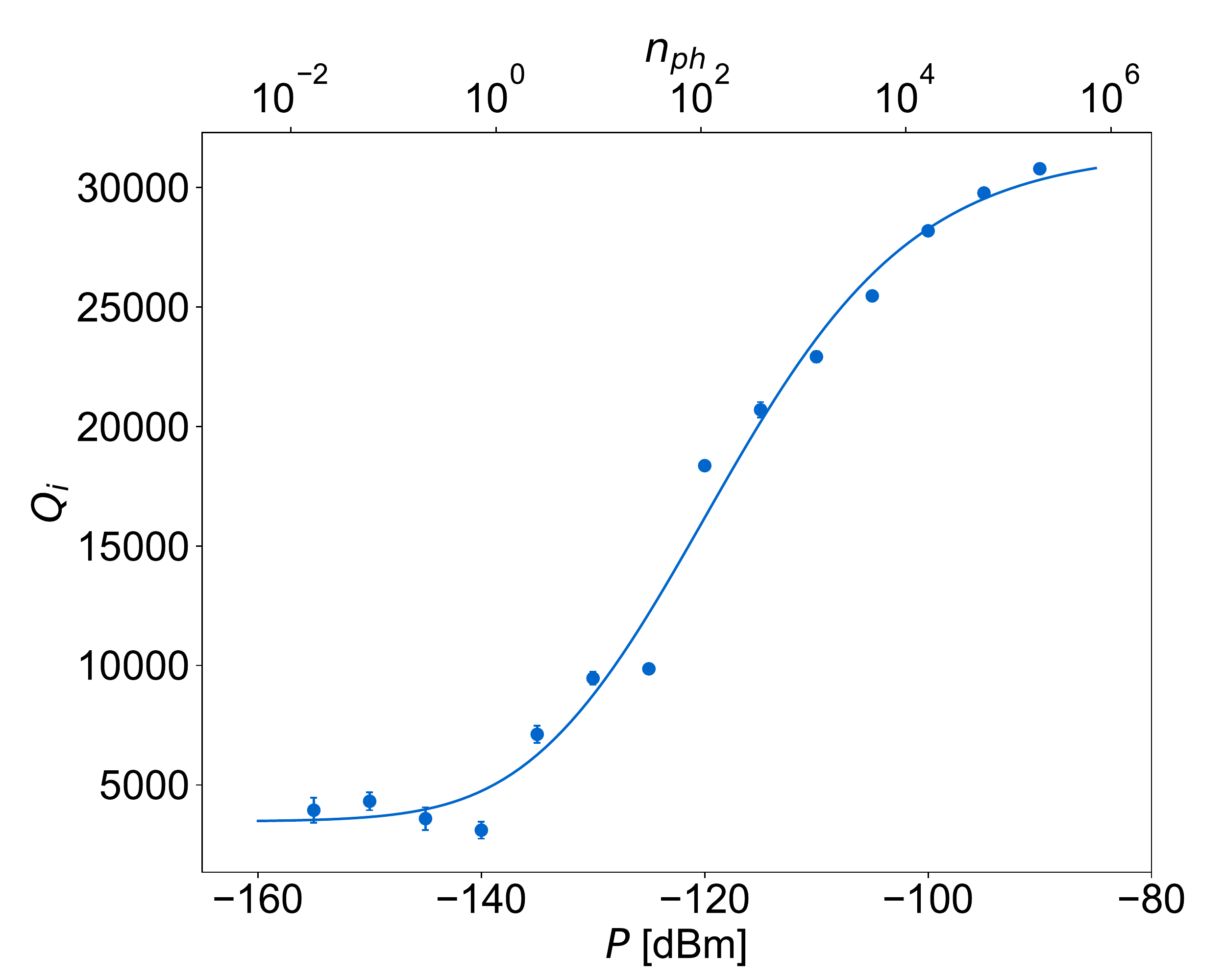}
    \caption{Power dependence of losses in the linear regime. Intrinsic quality factor versus power at $\xi/\xi_b \ll 1$ for the 0.1 $\mu$m wide KI NW resonator. The solid line is the fit using the TLS model.}
    \label{fig:TLS}
\end{figure}

We investigate the power dependence of the losses in the resonators in both the linear and nonlinear regimes. In the linear regime ($\xi/\xi_b \ll 1$), we observe losses behavior that agrees well with the model of the TLS-dominated resonator \cite{wang2009improving, sage2011study, sandberg2012etch} (Fig. \ref{fig:TLS}). At very low power, corresponding to less than single photon in the resonator, we get $Q_i \approx 4,000$. As the power increases $Q_i$ increases indicating the saturation of the TLS. We fit the data using the TLS model in Ref. \cite{wang2009improving}
\begin{equation}
    \frac{1}{Q_i} = \frac{1}{Q_0} + \frac{1}{Q_{\text{TLS}}}
\end{equation}
where $Q_0$ is the intrinsic quality factor when TLS is saturated and it is determined by other loss mechanisms, while $Q_{\text{TLS}}$ is given by
\begin{equation}
    Q_{\text{TLS}} = Q^{(0)}_{\text{TLS}} \sqrt{1+\left(\frac{P}{P_c}\right)^\beta}
\end{equation}
where $Q^{(0)}_{\text{TLS}}$ is the intrinsic quality factor for unsaturated TLS which limits $Q_i$ at low power, $P$ is the input power, $P_c$ is the critical power above which TLS saturation starts to become significant, and $\beta$ is a rescaling factor of the order of 1. The measurement of $Q_i$ versus power of the 0.1 $\mu$m wide resonator is fitted to this expression and best fit is shown in Fig. \ref{fig:TLS}, with $Q^{(0)}_{\text{TLS}}=3,900$ and $Q_0=31,500$.

\begin{figure}
    \centering
    \includegraphics [width=\linewidth] {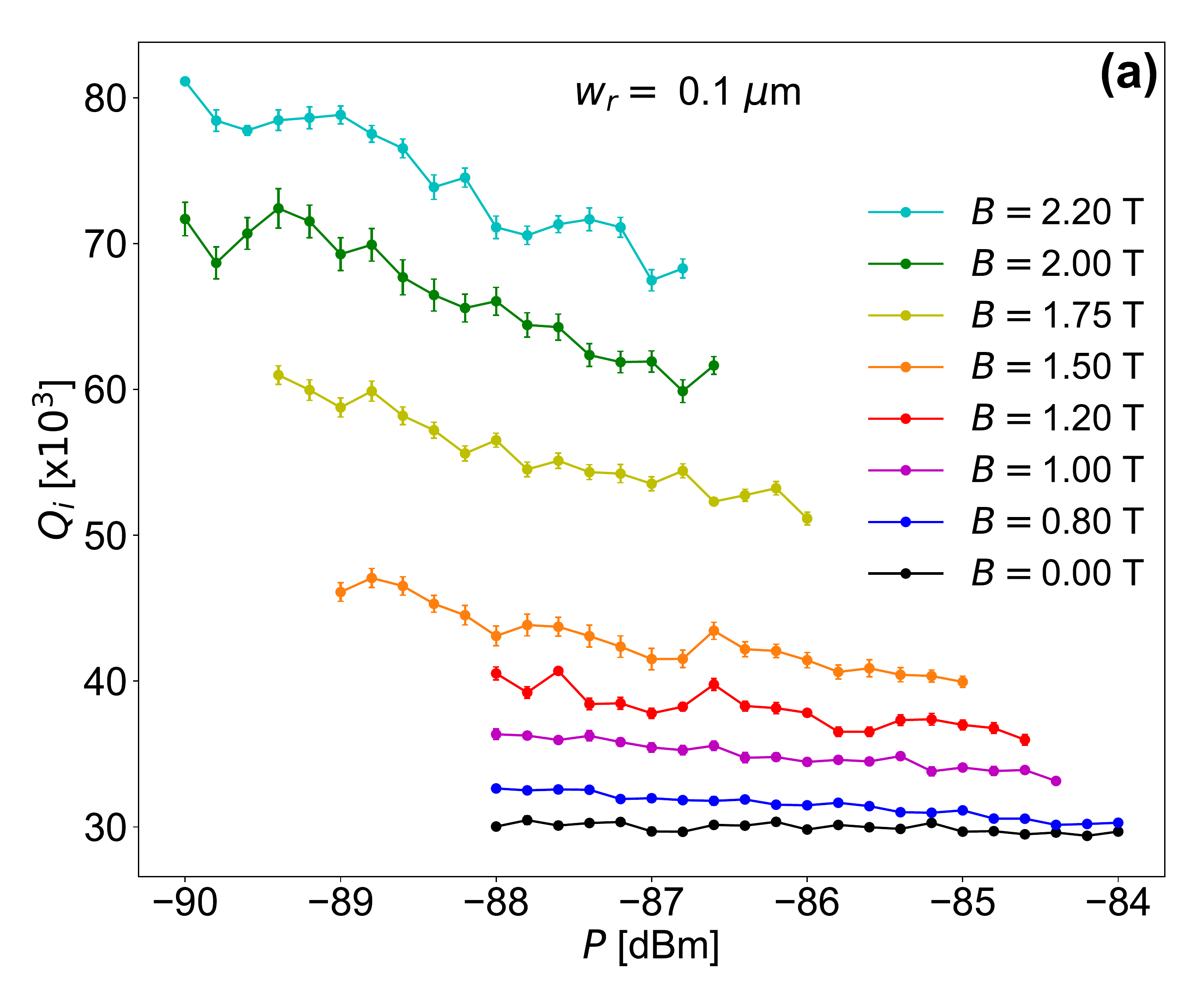}
    \includegraphics [width=\linewidth] {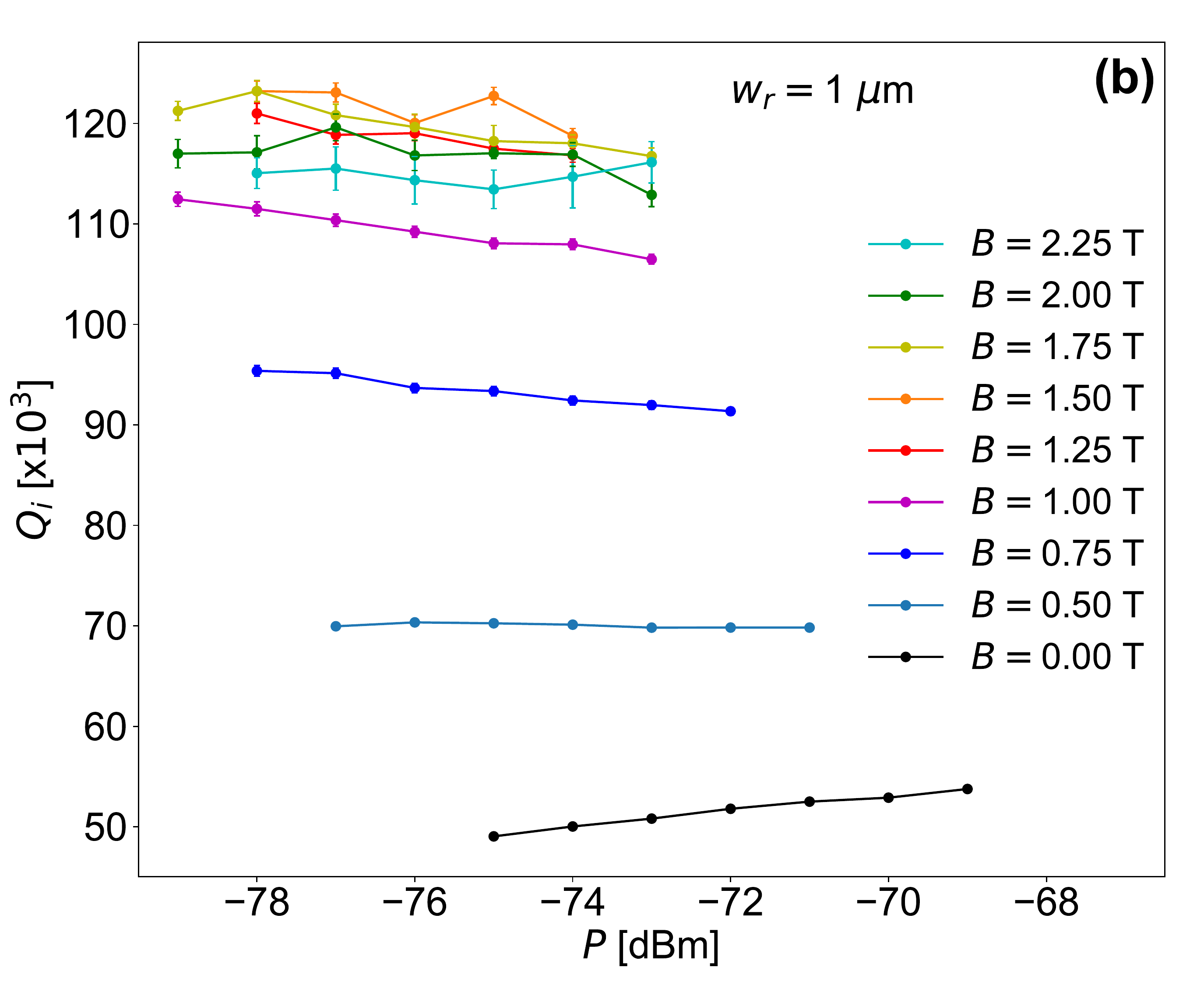}
    \caption{Power dependence of losses in the nonlinear regime. Intrinsic quality factor versus power at $0.1 < \xi/\xi_b < 1$ for the (a) 0.1 $\mu$m and (b) 1 $\mu$m wide KI NW resonators at different magnetic fields.}
    \label{fig:QiPB}
\end{figure}

The power dependence of $Q_i$ in the nonlinear regime ($0.1 < \xi/\xi_b < 1$) is shown in Fig. \ref{fig:QiPB} for resonators of widths 0.1 and 1 $\mu$m at different in-plane magnetic fields. For the 0.1 $\mu$m resonator (Fig. \ref{fig:QiPB}(a)), at zero magnetic field $Q_i$ is stable with power. As the magnetic field increases $Q_i$ increases in general, as illustrated in Fig. \ref{fig:Overall}(d), however it starts decreasing with increasing power. Higher magnetic field enhances the decrease of $Q_i$ versus power. There are two possible explanations for this behavior. The first one is that increasing power generates thermal quasi-particles, and as the magnetic field increases the scattering of these quasi-particles increases causing higher losses. The second possible explanation is that the magnetic field generates  vortices whose velocity depends on the driving power. As the power increases, the vortices acquire higher velocities, which increases the losses associated to them \cite{nsanzineza2016vortices}.

A similar behavior of increasing $Q_i$ with magnetic field and decreasing $Q_i$ with power at higher fields, is observed in the 1 $\mu$m NW resonator (Fig. \ref{fig:QiPB}(b)). However, here at zero magnetic field $Q_i$ is increasing with power, indicating that the power is not high enough to fully saturate the TLS in this resonator. Another interesting observation in the 1 $\mu$m resonator is that $Q_i$ increases rapidly with magnetic field compared to the 0.1 $\mu$m resonator, such that $Q_i$ reaches its maximum value at $B\approx1.5$ T. This is probably due to the coupling of the 1 $\mu$m resonator to a large number of magnetic centers in the substrate because of its large dimensions (length and width) compared to the 0.1 $\mu$m resonator.

\section{Magnetic Field Effect on Nonlinearity}
\label{app:KVsB}
Although vortex generation is strongly suppressed for in-plane magnetic field thanks to the small thickness of the film, the magnetic field still changes $T_c$ due to breaking the time-reversal degeneracy of the Cooper pairs inducing an effective depairing energy $E_d$. In the dirty limit, where the mean-free-path, limited by the film thickness $t=10$ nm, is smaller than London penetration depth $\sim200$ $\mu$m for NbTiN \cite{valente2016superconducting}, $\Delta T_c$ scales linearly with $E_d$ according to $k_B \Delta T_c = -(\pi/4)E_d/2$. For in-plane magnetic field $B$, $E_d$ can be approximated by $E_d = D e^2 B^2 t^2 / 3\hbar$ \cite{tinkham2004introduction}. Therefore $\Delta T_c$ scales quadratically with $B$ such that
\begin{equation}
    \Delta T_c(B) = - \frac{\pi}{24} \frac{D e^2 t^2}{\hbar k_B} B^2.
\end{equation}
This change in $T_c$ results in a shift in the kinetic inductance and the critical current due to their dependence on the superconducting gap. In the case of weak nonlinearity ($I \ll I_c$) and at $T \ll T_c$, $L_k \propto 1/T_c$ and $I_c \propto T_c^{3/2}$ \cite{annunziata2010tunable}. Therefore $\Delta L_k / L_k = - \Delta T_c / T_c$ and $\Delta I_c / I_c = (3/2) \Delta T_c / T_c$.

To investigate how $\Delta T_c(B)$ affects the resonator nonlinearity, we write down the expression for the self-Kerr coefficient of a KI resonator as \cite{parker2021near}
\begin{equation}
    K_0 \propto -\frac{\hbar \omega_0}{ L_k I_c^2} \omega_0 = -\frac{\hbar}{4l_r C L_k^2 I_c^2},
\end{equation}
where $l_r$ is the length of the resonator, $\omega_0 = 1/2l_r\sqrt{LC}$ is the resonance frequency, $L$ is the total inductance per unit length, which is dominated by the kinetic inductance ($L \approx L_k/l_r$), and $C$ is the capacitance per unit length. The total change in $K_0$ is then given by
\begin{equation}\label{eqn:Del_K}
    \frac{\Delta K_0}{K_0} = - \frac{\Delta T_c}{T_c} = \frac{\pi}{24} \frac{D e^2 t^2}{\hbar k_B T_c} B^2.
\end{equation}
According to (\ref{eqn:Del_K}) a magnetic field of 2 T is expected to cause an increase of only $0.1\%$ in $|K_0|$, which is beyond our measurement accuracy.


\providecommand{\noopsort}[1]{}\providecommand{\singleletter}[1]{#1}%

\end{document}